\documentclass[sigconf,nonacm]{acmart}

\AtBeginDocument{%
  \providecommand\BibTeX{{%
    \normalfont B\kern-0.5em{\scshape i\kern-0.25em b}\kern-0.8em\TeX}}}

\usepackage{enumitem}
\usepackage{epsfig}
\usepackage{graphicx}
\usepackage{algorithm}
\usepackage{algorithmicx}
\usepackage{algpseudocode}
\usepackage{amsmath}
\usepackage{listings}
\usepackage{xspace}
\usepackage{caption}
\usepackage{subcaption}
\usepackage{tabularx}
\usepackage{longtable}

\usepackage{multirow}
\usepackage{booktabs}
\usepackage{tikz}
\usepackage[most]{tcolorbox}

\usepackage{soul}

\def\BibTeX{{\rm B\kern-.05em{\sc i\kern-.025em b}\kern-.08em
    T\kern-.1667em\lower.7ex\hbox{E}\kern-.125emX}}

\renewcommand{\em}{\it}

\newcommand{\ignore}[1]{}

\def\cfigure[#1,#2,#3]{
\begin{figure}
\begin{center}

\includegraphics[width=3.2in]{#1} 
 
\caption[]{#2
} \label{#3}
 
\end{center}
\vspace*{-0.3in}
\end{figure}}

\def\cfiguredouble[#1,#2,#3,#4]{
\begin{figure}
\begin{center}
\includegraphics[width=3in]{#1} 
\\(a)\\
\includegraphics[width=3in]{#2} 
\\(b)\\
\caption[]{#3} \label{#4}
\end{center}
\vspace*{-0.2in}
\end{figure}}

\def\cfiguretemp[#1,#2,#3]{
\begin{figure}
\vspace*{0mm}
\begin{center}

\includegraphics[width=3.5in]{#1} 
 
\vspace*{-3mm}\caption[]{#2
} \label{#3}
 
\vspace*{-5mm}
\end{center}
\vspace*{-2mm}
\end{figure}}

\def\wfigure[#1,#2,#3]{
\begin{figure*}[t]
\vspace*{0mm}
\begin{center}

\includegraphics[width=6.4in]{#1} 
 
\vspace*{-3mm}\caption[]{#2
} \label{#3}
 
\vspace*{-5mm}
\end{center}
\vspace*{-2mm}
\end{figure*}}

\def\threefigure[#1,#2,#3,#4,#5]{
\begin{figure*}
\vspace*{0mm}
\begin{center}

\begin{tabular}{ccc}
\includegraphics[width=2in]{#1} & \includegraphics[width=2in]{#2} &  \includegraphics[width=2in]{#3} \\
(a) & (b) & (c) \\
\end{tabular}

\vspace*{-3mm}\caption[]{#4
} \label{#5}

\vspace*{-5mm}
\end{center}
\vspace*{-2mm}
\end{figure*}}

\def\threefigureSC[#1,#2,#3,#4,#5]{
\begin{figure}
\vspace*{0mm}
\begin{center}

\begin{tabular}{ccc}
\hspace{-0.1in}\includegraphics[width=1.2in]{#1} &
\hspace{-0.15in}\includegraphics[width=1.2in]{#2} &
\hspace{-0.15in}\includegraphics[width=1.2in]{#3} \\
(a) & (b) & (c) \\
\end{tabular}
\caption[]{#4
} \label{#5}
\vspace*{-0.1in}
\end{center}
\vspace*{-0.2in}
\end{figure}}

\def\dcfigure[#1,#2,#3,#4,#5,#6]{
{
\begin{figure*}
\vspace*{0.2in}\
\begin{center}
\begin{minipage}[c]{3in}{
\includegraphics[width=3in]{#1} 
\vspace*{-3mm}\caption[]{#2} \label{#3} \
}\end{minipage}\hspace*{0.5in}\
\begin{minipage}[c]{3in}{
\includegraphics[width=3in]{#4} 
\vspace*{-3mm}\caption[]{#5}\label{#6} \
}\end{minipage}
\end{center}
\vspace*{-0.4in}\
\end{figure*}
}
}

\def\qcfigure[#1,#2,#3,#4,#5,#6]{
{
\begin{figure*}
\vspace*{0.2in}\
\begin{center}
\begin{minipage}[c]{3in}{
\includegraphics[width=3in]{#1} 
\vspace*{-3mm}
}
\end{minipage}\hspace*{0.5in}\
\begin{minipage}[c]{3in}{
\includegraphics[width=3in]{#2} 
\vspace*{-3mm}
}\end{minipage}

\begin{minipage}[c]{3in}{
\includegraphics[width=3in]{#3} 
\vspace*{-3mm}
}
\end{minipage}\hspace*{0.5in}\
\begin{minipage}[c]{3in}{
\includegraphics[width=3in]{#4} 
\vspace*{-3mm}
}\end{minipage}
\end{center}
\caption[]{#5}\label{#6}
\end{figure*}
}
}

\def\twfigureabc[#1,#2,#3,#4,#5]{
{
\tiny
\begin{figure}
\begin{center}
\begin{minipage}[c]{3.4in}{
\includegraphics[width=3.4in]{#1} 
\vspace*{-4mm}
}
\end{minipage}
\\(a)\\

\begin{minipage}[c]{3.4in}{
\includegraphics[width=3.4in]{#2} 
\vspace*{-4mm}
}\end{minipage}
\\(b)\\

\begin{minipage}[c]{3.4in}{
\includegraphics[width=3.4in]{#3} 
\vspace*{-4mm}
}
\end{minipage}
\\(c)\\
\end{center}
\vspace*{-6mm}
\caption[]{#4}
\label{#5}
\end{figure}
}
}

\def\twfigure[#1,#2,#3,#4,#5]{
{
\begin{figure}
\vspace*{0.2in}\
\begin{center}
\begin{minipage}[c]{6.5in}{
\includegraphics[width=6.5in]{#1} 
\vspace*{-3mm}
}
\end{minipage}

\begin{minipage}[c]{6.5in}{
\includegraphics[width=6.5in]{#2} 
\vspace*{-3mm}
}\end{minipage}

\begin{minipage}[c]{6.5in}{
\includegraphics[width=6.5in]{#3} 
\vspace*{-3mm}
}
\end{minipage}
\end{center}
\caption[]{#4}\label{#5}
\end{figure}
}
}

\def\dwfigure[#1,#2,#3,#4]{
{
\begin{figure*}
\vspace*{0.2in}\
\begin{center}
\begin{minipage}[c]{3.5in}{
\includegraphics[width=3.5in]{#1} 
\vspace*{-3mm}
}
\end{minipage}

\begin{minipage}[c]{3.5in}{
\includegraphics[width=3.5in]{#2} 
\vspace*{-3mm}
}\end{minipage}

\end{center}
\caption[]{#3}\label{#4}
\end{figure*}
}
}

\def\dssfigure[#1,#2,#3,#4,#5,#6]{
{
\begin{figure*}
\vspace*{0.2in}\
\begin{center}
\begin{minipage}[c]{4in}{
\includegraphics[width=4in]{#1}
\vspace*{-3mm}\caption[]{#2} \label{#3} \
}\end{minipage}\hspace*{0.5in}\
\begin{minipage}[c]{2in}{
\includegraphics[width=2in]{#4}
\vspace*{-3mm}\caption[]{#5}\label{#6} \
}\end{minipage}
\end{center}
\vspace*{-0.4in}\
\end{figure*}
}
}

\def\dsfigure[#1,#2,#3,#4,#5,#6]{
{
\begin{figure*}
\vspace*{0.2in}\
\begin{center}
\begin{minipage}[c]{3in}{
\includegraphics[width=3in]{#1}
\vspace*{-3mm}\caption[]{#2} \label{#3} \
}\end{minipage}\hspace*{0.5in}\
\begin{minipage}[c]{3in}{
\hspace*{0.5in}\
\includegraphics[height=3in]{#4}
\vspace*{-3mm}\caption[]{#5}\label{#6} \
}\end{minipage}
\end{center}
\vspace*{-0.4in}\
\end{figure*}
}
}

\def\dsyfigure[#1,#2,#3,#4,#5,#6]{
{
\begin{figure*}
\vspace*{0.2in}\
\begin{center}
\begin{minipage}[c]{2.5in}{
\includegraphics[height=2.5in]{#1}
\vspace*{-3mm}\caption[]{#2} \label{#3} \
}\end{minipage}\hspace*{0.5in}\
\begin{minipage}[c]{2.5in}{
\includegraphics[height=2.5in]{#4}
\vspace*{-3mm}\caption[]{#5}\label{#6} \
}\end{minipage}
\end{center}
\vspace*{-0.4in}\
\end{figure*}
}
}

\def\dyfigure[#1,#2,#3,#4,#5,#6]{
{
\begin{figure*}
\vspace*{0.2in}\
\begin{center}
\begin{minipage}[c]{3in}{
\includegraphics[height=3in]{#1} 
\vspace*{-3mm}\caption[]{#2} \label{#3} \
}\end{minipage}\hspace*{0.5in}\
\begin{minipage}[c]{3in}{
\includegraphics[height=3in]{#4} 
\vspace*{-3mm}\caption[]{#5}\label{#6} \
}\end{minipage}
\end{center}
\vspace*{-0.4in}\
\end{figure*}
}
}

\def\dyoldfigure[#1,#2,#3,#4,#5,#6]{
{
\begin{figure*}
\vspace*{0.2in}\
\begin{center}
\begin{minipage}[c]{3in}{
\epsfysize=2.0in\
\hspace{0.5in}\
\epsfbox{#1}
\vspace*{-3mm}\caption[]{#2} \label{#3} \
}\end{minipage}\hspace*{0.25in}\
\begin{minipage}[c]{3in}{
\epsfysize=2.0in\
\hspace{0.5in}\
\epsfbox{#4}
\vspace*{-3mm}\caption[]{#5}\label{#6} \
}\end{minipage}
\end{center}
\vspace*{-0.4in}\
\end{figure*}
}
}

\def\wpfigure[#1,#2,#3,#4]{
\begin{figure*}
\vspace*{4mm}
\begin{center}

\includegraphics[width=#4]{#1} 

\vspace*{-3mm}\caption[]{#2
} \label{#3}

\vspace*{-5mm}
\end{center}
\end{figure*}}

\def\wprfigure[#1,#2,#3,#4,#5]{
\begin{figure*}
\vspace*{4mm}
\begin{center}

\includegraphics[width=#4, angle=#5]{#1} 

\vspace*{-3mm}\caption[]{#2
} \label{#3}

\vspace*{-5mm}
\end{center}
\end{figure*}}

\def\DoubleFigureWSlide[#1,#2,#3,#4,#5,#6,#7,#8,#9]{
\begin{figure*}
\vspace*{#9}
\begin{center}
\begin{minipage}{#4}
\includegraphics[width=#4]{#1}
\vspace*{-3mm}\caption{#2
}\label{#3}
\end{minipage}
\hspace{2em}
\begin{minipage}{#8}
\includegraphics[width=#8]{#5}
\vspace*{-3mm}\caption{#6
}\label{#7}
\end{minipage}
\vspace*{-5mm}
\end{center}
\end{figure*}
}

\def\DoubleFigureW[#1,#2,#3,#4,#5,#6,#7,#8]{
\begin{figure*}
\vspace*{0in}
\begin{center}
\begin{minipage}{#4}
\includegraphics[width=#4]{#1}
\vspace*{-3mm}\caption{#2
}\label{#3}
\end{minipage}
\hspace{2em}
\begin{minipage}{#8}
\includegraphics[width=#8]{#5}
\vspace*{-3mm}\caption{#6
}\label{#7}
\end{minipage}
\vspace*{-5mm}
\end{center}
\end{figure*}
}

\def\DoubleFigureWHack[#1,#2,#3,#4,#5,#6,#7,#8]{
\begin{figure*}
\vspace*{0in}
\begin{center}
\begin{minipage}{3in}
\includegraphics[width=#4]{#1}
\vspace*{-3mm}\caption{#2
}\label{#3}
\end{minipage}
\hspace{2em}
\begin{minipage}{3in}
\includegraphics[width=#8]{#5}
\vspace*{-3mm}\caption{#6
}\label{#7}
\end{minipage}
\vspace*{-5mm}
\end{center}
\end{figure*}
}

\def\ddcfigure[#1,#2,#3,#4]{
\begin{figure*}[t]
\vspace*{0.2in}\
\begin{center}
\begin{minipage}[c]{3in}{
\includegraphics[width=3in]{#1} 
}\end{minipage}\hspace{0.5in}\
\begin{minipage}[c]{3in}{
\includegraphics[width=3in]{#2} 
}\end{minipage}\vspace*{-0.10in} \caption[]{#3}\label{#4}
\end{center}
\vspace*{-0.4in}\
\end{figure*}
}

\def\dddcfigure[#1,#2,#3,#4]{
\begin{figure*}
\vspace*{0.2in}\
\begin{center}
\begin{tabular}{cc}
\includegraphics[width=3in]{#1} &
\includegraphics[width=3in]{#2} \\
(a) & (b) \\
\end{tabular}\vspace*{-0.10in}\caption[]{#3}\label{#4}
\end{center}
\vspace*{-0.4in}\
\end{figure*}
}

\def\qqcfigure[#1,#2,#3,#4,#5,#6]{
\begin{figure*}[t]
\begin{center}
\begin{tabular}{cc}
\includegraphics[width=3.3in]{#1} &
\includegraphics[width=3.3in]{#2} \\
(a) & (b) \\
\includegraphics[width=3.3in]{#3} &
\includegraphics[width=3.3in]{#4} \\
(c) & (d) \\
\end{tabular}\vspace*{-0.10in}\caption[]{#5}\label{#6}
\end{center}
\vspace*{-0.4in}\
\end{figure*}
}
\def\qqcfiguresinglecol[#1,#2,#3,#4,#5,#6]{
\begin{figure}[t]
\begin{center}
\begin{tabular}{cc}
\hspace*{-0.2in}\includegraphics[width=1.8in]{#1} &
\hspace*{-0.2in}\includegraphics[width=1.8in]{#2} \\
\hspace*{-0.2in}(a) & 
\hspace*{-0.2in}(b) \\
\hspace*{-0.2in}\includegraphics[width=1.8in]{#3} &
\hspace*{-0.2in}\includegraphics[width=1.8in]{#4} \\
\hspace*{-0.2in}(c) & 
\hspace*{-0.2in}(d) \\
\end{tabular}\vspace*{-0.10in}\caption[]{#5}\label{#6}
\end{center}
\vspace*{-0.4in}\
\end{figure}
}

\def\ddcfiguresinglecol[#1,#2,#3,#4]{
\begin{figure}[t]
\begin{center}
\begin{tabular}{cc}
\hspace*{-0.1in}\includegraphics[width=1.8in]{#1} &
\hspace*{-0.1in}\includegraphics[width=1.8in]{#2} \\
\hspace*{-0.1in}(a) & 
\hspace*{-0.1in}(b) \\
\end{tabular}\vspace*{-0.10in}\caption[]{#3}\label{#4}
\end{center}
\vspace*{-0.4in}\
\end{figure}
}

\def\qqcfigureInAColumn[#1,#2,#3,#4,#5,#6]{
\begin{figure}[t]
\begin{center}
\includegraphics[width=3.3in]{#1} \\
(a)\\
\includegraphics[width=3.3in]{#2} \\
(b) \\
\includegraphics[width=3.3in]{#3} \\
(c) \\
\includegraphics[width=3.3in]{#4} \\
(d) \\
\vspace*{-0.10in}\caption[]{#5}\label{#6}
\end{center}
\vspace*{-0.5in}\
\end{figure}
}

\def\sixfigure[#1,#2,#3,#4,#5,#6,#7,#8]{
\begin{figure*}[t]
\vspace*{0.2in}\
\begin{center}
\begin{tabular}{cc}
\includegraphics[width=3in]{#1} &
\includegraphics[width=3in]{#2} \\
(a) & (b) \\
\includegraphics[width=3in]{#3} &
\includegraphics[width=3in]{#4} \\
(c) & (d) \\
\includegraphics[width=3in]{#5} &
\includegraphics[width=3in]{#6} \\
(e) & (f) \\
\end{tabular}\vspace*{-0.10in}\caption[]{#7}\label{#8}
\end{center}
\vspace*{-0.4in}\
\end{figure*}
}

\def\ddcfigureSlide[#1,#2,#3,#4,#5]{
\begin{figure*}
\vspace*{#5}\
\begin{center}
\begin{minipage}[c]{3in}{
\includegraphics[height=3in]{#1} 
}\end{minipage}\hspace{0.5in}\
\begin{minipage}[c]{3in}{
\includegraphics[height=3in]{#2} 
}\end{minipage}\vspace*{-0.10in} \caption[]{#3}\label{#4}
\end{center}
\vspace*{-0.4in}\
\end{figure*}
}

\def\dcfigureSingleCol[#1,#2,#3,#4]{
\begin{figure}
\begin{center}
\begin{tabular}{cc}
\hspace*{-0.2in}\includegraphics[width=1.5in]{#1} &
\hspace*{-0.2in}\includegraphics[width=1.5in]{#2} \\
\hspace*{-0.2in}(a) & 
\hspace*{-0.2in}(b) \\
\end{tabular}\vspace*{-0.10in}\caption[]{#3}\label{#4}
\end{center}
\vspace*{-0.4in}\
\end{figure}
}

\def\cxfigure[#1,#2,#3]{
\begin{figure}
\vspace*{4mm}
\begin{center}
 
\epsfxsize=2.5in\
\epsfbox{#1}\
 
\vspace*{-0.10in}\caption[]{#2
} \label{#3}
 
\vspace*{-5mm}
\end{center}
\vspace*{-2mm}
\end{figure}}

\newcommand{\x}{$\times$}

\newif\ifremark
\long\def\remark#1{
\ifremark%
        \begingroup%
        \dimen0=\columnwidth
        \advance\dimen0 by -1in%
        \setbox0=\hbox{\parbox[b]{\dimen0}{\protect\em #1}}
        \dimen1=\ht0\advance\dimen1 by 2pt%
        \dimen2=\dp0\advance\dimen2 by 2pt%
        \vskip 0.25pt%
        \hbox to \columnwidth{%
                \vrule height\dimen1 width 3pt depth\dimen2%
                \hss\copy0\hss%
                \vrule height\dimen1 width 3pt depth\dimen2%
        }%
        \endgroup%
\fi}

\remarktrue
\newcommand*\circled[1]{\tikz[baseline=(char.base)]{
            \node[shape=circle,fill,inner sep=0pt] (char)
{\textcolor{white}{#1}};}}

\begin{document}



\newcommand{\poan}[1]{\textcolor{blue}{[Po-An: #1]}}
\newcommand{\tmp}[1]{\textcolor{red}{#1}}

\lstdefinestyle{customc}{
  belowcaptionskip=1\baselineskip,
  breaklines=true,
  frame=L,
  xleftmargin=\parindent,
  language=C,
  showstringspaces=false,
  basicstyle=\footnotesize\ttfamily,
  keywordstyle=\bfseries\color{green!40!black},
  commentstyle=\itshape\color{purple!40!black},
  identifierstyle=\color[HTML]{0920C7},
  stringstyle=\color{orange},
}
\lstset{emph={half%
    },emphstyle={\color{green!40!black}\bfseries}%
}%

\lstdefinestyle{customasm}{
  belowcaptionskip=1\baselineskip,
  frame=L,
  xleftmargin=\parindent,
  language=[x86masm]Assembler,
  basicstyle=\footnotesize\ttfamily,
  commentstyle=\itshape\color{purple!40!black},
}

\lstset{escapechar=@,style=customc}

\newcommand{\mm}{mm$^2$}
\newcommand{\figtitle}[1]{\textbf{#1}}
\newcommand{\us}{$\mu$s}
\newcommand{\fixme}[1]{#1}
\newcommand{\adrian}[1]{{\color{green}\textbf{#1}}}
\newcommand{\laura}[1]{{\color{pink}\textbf{#1}}}
\newcommand{\joel}[1]{{\color{red}\textbf{#1}}}
\newcommand{\ameen}[1]{{\color{blue}\textbf{#1}}}
\newcommand{\arup}[1]{{\color{yellow}\textbf{#1}}}
\newcommand{\hungwei}[1]{{{#1}}}
\newcommand{\andrew}[1]{{{#1}}}
\newcommand{\boram}[1]{{\color{red}\textbf{#1}}}
\newcommand{\Bella}[1]{{\color{blue}\textbf{\textit{#1}}}}

\newcommand{\note}[2]{{\color{red}\fixme{$\ll$ #1 $\gg$ #2}}}
\newcommand{\myitem}[1]{\hspace*{-\parindent}\textbf{#1}\hspace*{\parindent}}

\newcommand{\LHEDB}{NSHEDB}
\newcommand{\TITLE}{NSHEDB}
\newcommand{\BASELINE}{HE$^3$DB}
\soulregister{\cite}{7}
\soulregister{\ref}{7}
\soulregister{\url}{7}
\soulregister{\hl}{1}

\newcommand{\hlA}[1]{\sethlcolor{yellow}\hl{#1}}
\newcommand{\hlB}[1]{\sethlcolor{green}\hl{#1}}
\newcommand{\hlC}[1]{\sethlcolor{cyan}\hl{#1}}
\newcommand{\reviewed}[1]{{{#1}}}

\newcommand{\shorttitle}{Revision Response}

\title{\LHEDB{}: Noise-Sensitive Homomorphic Encrypted Database Query Engine\textsuperscript{*}}
\author{Boram Jung}
\affiliation{%
  \institution{UC Riverside}
  \country{United States}
}
\email{bjung022@ucr.edu}

\author{Yuliang Li}
\orcid{0000-0002-1825-0097}
\affiliation{%
  \institution{Meta}
  \country{United States}
}
\email{yuliangli@meta.com}

\author{Hung-Wei Tseng}
\affiliation{%
  \institution{UC Riverside}
  \country{United States}
}
\email{htseng@ucr.edu}
\begin{abstract}



Homomorphic encryption (HE) enables computations directly on encrypted data, offering strong cryptographic guarantees for secure and privacy-preserving data storage and query execution. However, despite its theoretical power, practical adoption of HE in database systems remains limited due to extreme ciphertext expansion, memory overhead, and the computational cost of bootstrapping, which resets noise levels for correctness.

This paper presents \LHEDB{}\footnote{\hlC{Implementation available at \url{https://anonymous.4open.science/r/ShaftDB-242C}}}, a secure query processing engine designed to address these challenges at the system architecture level. \LHEDB{} uses word-level leveled HE (LHE) based on the BFV scheme to minimize ciphertext expansion and avoid costly bootstrapping. It introduces novel techniques for executing equality, range, and aggregation operations using purely homomorphic computation, without transciphering between different HE schemes (e.g., CKKS/BFV$\leftrightarrow$TFHE) or relying on trusted hardware. Additionally, it incorporates a noise-aware query planner to extend computation depth while preserving security guarantees.

We implement and evaluate \LHEDB{} on real-world database workloads (TPC-H) and show that it achieves 20\x{}–1370\x{} speedup and a 73\x{} storage reduction compared to state-of-the-art HE-based systems, while upholding 128-bit security in a semi-honest model with no key release or trusted components.

\end{abstract}

\maketitle
\let\svthefootnote\thefootnote
\let\thefootnote\relax\footnotetext{\textsuperscript{*}This paper will appear in the 2026 ACM SIGMOD/PODS Conference.}
\addtocounter{footnote}{-1}\let\thefootnote\svthefootnote
\section{Introduction}
\label{sec:introduction}
Homomorphic encryption (HE) offers a compelling foundation for building privacy-preserving database systems, enabling secure computation over encrypted data without revealing sensitive contents. It provides end-to-end data confidentiality in untrusted environments, eliminating the need for trusted execution environments (TEEs) or key escrow. However, despite this strong cryptographic guarantee, practical deployment of HE-based database systems remains elusive due to severe performance, storage, and usability constraints.

\begin{itemize}
\item \textbf{Performance:} Modern HE operations depend on a time-consuming bootstrapping process to reset the noise level in computed data, ensuring the correctness of subsequent computations. Existing HE-based database systems also rely on costly transciphering between various HE schemes to support a wider range of database operations. These systems take at least several minutes to perform queries~\cite{ren2022heda,he3db,zhang2024arcedb}.
\item \textbf{Memory and Storage:} HE schemes result in significant expansion in ciphertext volume as the schemes need to encode difficult-to-decrypt noises while preserving the budget for correct numerical computation. State-of-the-art HE libraries like OpenFHE~\cite{OpenFHE} have to expand data volume by 10,000$\times$ to support bit-level operations and meet the standard 128 security level. Storing HE-encrypted data on storage devices or main memory during runtime is economically unfeasible.
\item \textbf{Functionality:} Although system designers can optionally reduce bootstrapping operations or coarsen the granularity of encoding to decrease the computation and space overhead, each compromise would sacrifice the functionality of the resulting system.
\end{itemize}

This paper introduces \LHEDB{}, Noise-Sensitive Homomorphically Encrypted Database, \reviewed{a secure query processing engine designed to make HE-based encrypted databases viable in untrusted execution settings. Rather than proposing new cryptographic primitives, \LHEDB{} focuses on addressing the system-level bottlenecks that have hindered the practicality of HE-backed query execution. It achieves this through a combination of optimized data encoding, HE-friendly query transformations, and noise-aware execution planning—all while preserving strict security guarantees under a well-defined semi-honest adversary model. \LHEDB{} is not simply a performance optimization of previous HE DBMS designs—it is a secure system architecture that makes principled design decisions to minimize leakage, avoid scheme-switching transciphering, and operate within the constraints of leveled HE.}

The very first design decision of \LHEDB{} is to utilize the Leveled HE (LHE)~\cite{bgv14} mechanism, based on the \emph{Brakerski/Fan-Vercauteren (BFV) scheme}~\cite{bv14}, and to encrypt data at the \emph{word level}.
The use of LHE at the word level offers several advantages to \LHEDB{} compared to traditional HE-based database systems. First, this design choice allows \LHEDB{} to significantly reduce the data expansion rate to only 28\x{} relative to raw data, making storage usage more economical and preserving valuable interconnect bandwidth. Second, by encrypting data at the word level, \LHEDB{} leverages batch encoding, where multiple data points are encrypted into a single ciphertext, unlocking data-level parallelism to further enhance performance.
Finally, since LHE can execute a limited number of operations without bootstrapping, \LHEDB{} effectively circumvents the computational overhead in time-consuming bootstrapping operations. \reviewed{This makes query execution feasible within the computational budget of leveled HE, avoiding the need for expensive bootstrapping or compromise on security.} While approximate encryption methods like CKKS~\cite{ckks17} also offer large plaintext space and batch encoding, they are unsuitable for databases requiring exact query selection, as they rely on approximate arithmetic. Since exact comparisons are essential for database queries, BFV is a more suitable choice for \LHEDB{}.

However, simply implementing word-level LHE does not enable \LHEDB{} to perform database operations due to the lack of support for bitwise logical operations and arithmetic operations such as division. To address this, \LHEDB{} proposes a set of algorithms designed to efficiently execute critical database functions, including equality checks, range queries, averages, etc., without the need to re-encrypt data into bit-level HE on the fly.
Another constraint of LHE is the limited number of operations it can perform. This paper has identified that the sequence in which operations are performed significantly impacts the growth of the noise level. Therefore, \LHEDB{} meticulously evaluates each query to minimize the need to reset the noise level of computation results.
We have implemented and evaluated \LHEDB{}, comparing its performance with existing HE-based DB prototypes. \LHEDB{} is 20\x{} to 1370\x{} faster in a set of TPC-H queries. Furthermore, \LHEDB{} only requires 1.4\% of the space compared to its counterpart, or 28\x{} relative to the raw data storage without encryption.

\noindent
In summary, \LHEDB{} advances encrypted database systems through the following contributions:

\begin{enumerate}
    \item \textbf{A novel query processing architecture for FHE constraints:} We identify how the constraints of word-level homomorphic encryption (HE) impose a new architectural objective that renders traditional index structures ineffective. Our novel scan-first database architecture embraces this constraint, pivoting the properties of HE into a strength by maximizing data-level parallelism
    
    \item \textbf{Implementations of all major relational and arithmetic operators under FHE:} We provide the first comprehensive set of database-aware arithmetic algorithms implementing all major SQL operations (joins, aggregations, filters) entirely within a single homomorphic encryption scheme, eliminating costly encryption scheme conversions (e.g., BFV $\leftrightarrow$ TFHE or CKKS $\leftrightarrow$ TFHE).
    
    \item \textbf{The first noise-aware query optimizer for encrypted databases:} Unlike traditional optimizers that minimize I/O or CPU, we introduce logical planning that treats multiplicative depth as the primary cost. Our novel optimizations---the \emph{Predicate Pull-Up} and \emph{Mask-Injection Tuning} rewrites---are fundamentally database transformations that restructure query plans to minimize noise and eliminate time-consuming bootstrapping

    \item \reviewed{\textbf{A real-world implementation and benchmark evaluation demonstrating feasibility:}} \LHEDB{} is rigorously evaluated using the \textit{TPC-H} benchmark, demonstrating that secure query processing with leveled HE is feasible under practical workloads and a well-defined threat model.

\end{enumerate}


\section{Challenges of using HE in Database Systems}
\label{sec:background}

\subsection{Homomorphic Encryption}
\label{sec:he}
Homomorphic Encryption (HE) is a state-of-the-art encryption technique that allows for the computation of encrypted data without decryption. HE ensures the confidentiality of sensitive information, as all inputs, including user inputs and data storage, remain encrypted throughout the computation process.

\begin{example}
The following example illustrates the basic concept behind HE using one-bit data.
Assume the encryption function in an HE system is given by:
\begin{equation}
\label{eq:enc}
E(m) = m + 2 r + as
\end{equation}
where $m$ is a one-bit raw data, $s$ represents the secret key, $r << |s|$ is a small random noise, and $a$ is a larger, random noise. If we have two data values $m_1$ and $m_2$, and we encrypt $m_1$ and $m_2$ with different noise but the same key, then we can derive $E(m_1) = m_1 + 2 r_1 + a_1s$ and $E(m_2) = m_2 + 2 r_2 + a_2s$.
If we directly perform $x = E(m_1) +E(m_2)$, then we can get another encrypted value
\begin{align*}
x &= E(m_1) +E(m_2) = (m_1 + m_2) + 2(r_1+r_2) + (a_1+a_2)s \\
  &= (m_1 + m_2) + 2 r' + a's .
\end{align*}
If someone has the secret key $s$, they can decrypt $x$ by modulus $s$ and $2$ as
\begin{align*}
D(x) &= [(m_1 + m_2) + 2r' + a's ] \mod s \mod 2 \\
     &= [(m_1 + m_2) + 2r'] \mod 2 = m_1 + m_2
\end{align*}
The same theory works for multiplications. By computing $y = E(m_1) \times E(m_2)$, we can get an encrypted value
\begin{align*}
y &= E(m_1) \times E(m_2) = m_1 m_2 + (a_1 E(m_2) + a_2 E(m_1))s \\
  &+ 2(m_1 r_2 + r_1 m_2 + 2 r_1 r_2) = m_1 m_2 + a''s + 2 r''
\end{align*}
With $s$, we can decrypt $y$ and get $m_1 m_2$ through the same decryption function.
\end{example}

Through the above example, we can make several observations. First, computing $x$ or $y$ does not require $s$, the secret key. Second, anyone received $E(m_1)$, $E(m_2)$, and $x$ cannot easily derive $m_1$, $m_2$ and $D(x)$ without $s$. Third, even with $s$, the system cannot decrypt the encrypted data if the $a'$ and $r'$ or $a''$ and $r''$ are too large. Finally, when we compute $x$ or $y$, the noise level in $x$ and $y$ grows as the computation aggregates or multiplies the noise from $E(m_1)$ and $E(m_2)$. 

While the first two observations highlight the capability to compute on encrypted data without decryption, thereby ensuring security, the latter two present a key challenge for HE systems: managing noise growth. Modern HE approaches, such as Fully Homomorphic Encryption (FHE) and Leveled Homomorphic Encryption (LHE), offer different strategies to address this issue.

\subsubsection{Fully Homomorphic Encryption and Leveled Homomorphic Encryption}
\label{sec:differentHEs}

\subsubsection*{Fully Homomorphic Encryption}
Homomorphic operations introduce noise into ciphertexts, and if this noise exceeds a threshold, it can corrupt the decrypted results, limiting the number of computations that can be performed. Fully Homomorphic Encryption (FHE) resolves this by using bootstrapping, a process that resets the noise level, allowing for unlimited computations. This capability makes FHE suitable for evaluating arbitrary circuits and supporting complex applications~\cite{gentry2009fully}.

\subsubsection*{Leveled Homomorphic Encryption}
Leveled Homomorphic Encryption (LHE) manages noise growth by limiting the number of allowed multiplications, known as computation levels (or multiplicative depth). Each multiplication contributes to noise accumulation, and if the total noise exceeds a threshold, decryption errors may occur. Multiplicative depth measures the longest path of sequential multiplications from input to output.

\reviewed{Unlike Fully Homomorphic Encryption (FHE), which resets noise via bootstrapping to enable unlimited computation, LHE operates efficiently within a fixed multiplicative depth, reducing bootstrapping overhead. However, when computations exceed the predefined depth, LHE can selectively apply bootstrapping to enable deeper computations. Schemes like BFV~\cite{bv14}, BGV~\cite{bgv14}, and CKKS~\cite{ckks17} define the maximum allowable depth based on encryption parameters. While LHE aims to minimize bootstrapping, complex queries may still leverage it selectively when necessary.}

\subsubsection{Multiplicative Depth}
In homomorphic encryption, ciphertexts accumulate \textit{noise} with each operation, and deeper computations risk exceeding the noise budget—the maximum noise a ciphertext can tolerate before decryption fails. Multiplicative depth captures the longest sequence of multiplications from input to output. Formally, for two ciphertexts $A$ and $B$ with depths $\mathrm{depth}(A)$ and $\mathrm{depth}(B)$, their product $C = A \times B$ satisfies:
\[
 \mathrm{depth}(C) = \max(\mathrm{depth}(A), \mathrm{depth}(B)) + 1.
\]
\reviewed{As multiplications add new noise and amplify existing noise, computations with high multiplicative depth increase the risk of surpassing the noise budget and bootstrapping. However, careful parameter selection in LHE can ensure computations remain within a manageable noise budget, reducing or eliminating the need for bootstrapping.

In addition to the multiplicative depth that provides a straightforward measure of noise growth, \emph{the order} of multiplications can potentially change the noise aggregation. Operations like \emph{relinearization} or \emph{rescaling} occur after each multiplication and can introduce additional noise, depending on the intermediate ciphertext states. Thus, even if two multiplication strategies ultimately implement the same expression (e.g., \(x^4\)) --- one as \(\,(x \times x) \times (x \times x)\) and another as \(\,x \times \bigl(x \times (x \times x)\bigr)\) --- they can exhibit different intermediate noise behavior, potentially causing one strategy to reach the noise budget sooner than the other.}

\subsubsection{Bootstrapping}
During homomorphic computations, noise accumulates in ciphertexts, and once it exceeds a threshold, it can corrupt decryption results. Bootstrapping resets the noise level, enabling continued computations without loss of accuracy. This technique applies the decryption algorithm homomorphically, reducing noise while keeping data encrypted. Bootstrapping is crucial in FHE for supporting unlimited computations. Although specifics may vary across HE schemes, the primary goal remains the same: to reset noise and ensure computational integrity.

\subsubsection{Data Encryption Granularity of HE}
\label{sec:granularity}
The granularity of data encryption significantly affects the efficiency and capabilities of HE schemes. Two primary types of encryption schemes that support exact computations are: (1) bit-level encryption (e.g., TFHE~\cite{tfhe20}, FHEW~\cite{fhew15}), and (2) word-level encryption (e.g., BGV~\cite{bgv14}, BFV~\cite{bv14}).

\paragraph{Bit-Level Encryption}
Bit-level encryption, based on Boolean circuits, operates on individual bits and supports logical operations like AND, OR, and XOR. It is useful for non-linear algorithms and comparison operations, but its efficiency is limited due to significant data expansion, as each ciphertext handles only a few bits of data, leading to high storage and computational costs.

\paragraph{Word-Level Encryption}
In contrast, word-level encryption operates on integers using arithmetic circuits. It offers better efficiency for linear algorithms and supports batch encoding, where multiple integers are packed into a single ciphertext. This enables data-level parallelism, making word-level encryption more space- and computationally-efficient. However, it is less effective for operations requiring fine-grained bitwise logic.

\subsubsection{Data Type Encoding in BFV}
\label{sec:data_type_encoding}
Since BFV operates on integers, non-integer data types require encoding. Floating-point numbers are encoded as fixed-point representations to maintain precision, while Boolean values map to integers (0 for False, 1 for True). For string data, we employ dictionary encoding, assigning sequential integer IDs to unique string values.

\subsubsection{BFV Operations}
\label{sec:bfv_ops}
The BFV scheme provides native support for arithmetic operations on packed ciphertexts. \texttt{Addition} and \texttt{Subtraction} operate element-wise between ciphertexts, as does \texttt{Multiplication}. \texttt{Negation} flips the sign of all packed values, while \texttt{Rotation} cyclically shifts elements by a specified offset.

Building on these primitives, HE systems implement two derived operations critical for database queries. \texttt{Extract} isolates a single element by multiplying with a basis vector (all zeros except a single 1 at the target position). \texttt{Broadcast} replicates a value across all slots via extraction followed by rotations and additions. These derived operations enable the fine-grained data manipulation required for query processing.
\subsubsection{Basic logical comparisons}
\label{sec:equality}
\paragraph{Equality}
HE systems commonly evaluate equality using Fermat's Little Theorem~\cite{wiles1995modular,kim2016better,gouert2023sok,iz20,cetin2015arithmetic}. While one could theoretically achieve bitwise comparisons in BFV by setting the plaintext modulus to 2, this fundamentally changes the data representation. Integers must be bit-decomposed across multiple slots (e.g., 16 slots for a 16-bit integer), reducing capacity by 16 times. Moreover, arithmetic operations would require complex bit-level circuits with rotations, eliminating BFV's efficient SIMD arithmetic advantages. Queries mixing bitwise and arithmetic operations would require maintaining dual encryption schemes with different moduli.

Therefore, arithmetic approaches using standard plaintext moduli are strongly preferred. These approaches leverage Fermat's Little Theorem to perform comparisons arithmetically. For prime $p$ and non-zero $a$, Fermat's Little Theorem states:
\begin{equation} 
    a^{(p-1)} \equiv 1 \pmod{p}
\end{equation}

This property enables equality testing: the expression $(x-y)^{(p-1)}$ equals 1 for non-zero differences and is undefined for zero. The system computes equality between $x$ and $y$ as:
\begin{equation} \label{eq:equality}
    EQ(x, y) = 1 - (x - y)^{(p-1)}
\end{equation}
This computation yields encrypted 1 when values are equal, 0 otherwise.

\paragraph{Range Comparisons}
Range comparisons extend the Fermat's Little Theorem approach by checking whether the difference $x - y$ falls within the negative range $[-\frac{p-1}{2}, -1]$. Under modular arithmetic, negative values map to the upper half of the field. The principle follows directly: $x < y$ holds if and only if $x - y$ yields a negative result. For instance, the comparison $3 < 5$ produces $3 - 5 = -2$, which falls in the negative range and thus evaluates to true, while $5 < 3$ produces $5 - 3 = 2$, which falls in the positive range and evaluates to false.

The system computes the less-than comparison by checking whether $x - y$ equals any value in the negative range:
\begin{equation} \label{eq:inequality}
LT(x,y) = \sum_{a=-\frac{p-1}{2}}^{-1} \left( 1 - (x-y-a)^{(p-1)} \right)
\end{equation}
This computation yields encrypted 1 when $x < y$, 0 otherwise. Systems can compute both equality and range comparisons simultaneously using the optimization from~\cite{iz20}.
\subsection{Challenges}
\label{sec:challenges}
Despite the properties of preserving the form of encrypted data during the whole computing process to ensure the privacy and security of user data, using HE in database systems is very challenging in the following aspects. 

\paragraph{Performance}
FHE performance characteristics create an entirely different optimization landscape for database systems. Bootstrapping---required to reset noise levels---dominates execution time~\cite{shokri2015privacy,de2021does,cousins2016fpga,roy2019fpga,samardzic2021f1,kim2022bts}. A breakdown of execution time for Boolean gate evaluations on 1-bit data using OpenFHE~\cite{OpenFHE} on our testbed shows that bootstrapping contributes to 99.99\% of the total execution time. Prior work~\cite{al2023demystifying} shows that CKKS takes 44 seconds to bootstrap 13-bit 32,768 elements (0.0013 seconds per bit), BGV takes 163 seconds to bootstrap 16-bit 1,024 elements (0.0099 seconds per bit). Recent OpenFHE implementations achieve TFHE bootstrapping at approximately 5ms per 1-bit gate on modern CPUs~\cite{OpenFHE}, with GPU and specialized hardware achieving even lower latencies. Recent studies on HE-based databases~\cite{zhang2024arcedb,he3db,ren2022heda} further highlight that bootstrapping dominates query processing time, accounting for up to 90\% of total execution time.
This fundamentally changes what database systems must optimize:
\begin{itemize}
\item Traditional cost: I/O and CPU cycles for data movement and computation
\item FHE cost: Multiplicative depth that accumulates noise toward bootstrapping thresholds
\item Traditional heuristics become counterproductive: predicate pushdown can trigger bootstrapping, turning a cheap filter into an expensive operation
\item Operation ordering dramatically affects total cost: the same query can vary from zero to multiple bootstraps depending on execution order
\end{itemize}
Database optimizers must be completely redesigned to minimize multiplicative depth rather than I/O patterns.

\paragraph{Space and Memory Efficiency}
Encrypting data significantly expands the size of the resulting ciphertexts. When using a state-of-the-art HE library that supports bit-level HE~\cite{OpenFHE} and meets the standard 128-bit security level, data expansion reaches 10,000$\times$. \reviewed{Similarly, HE-based databases~\cite{zhang2024arcedb,he3db,ren2022heda} that operate at the bit level experience data expansion of approximately 8,000$\times$, leading to substantial storage and computational overhead.} Storing data in bit-level FHE is economically infeasible. Even when assuming secure storage and converting data into HE form before offloading it to a third party, the ciphertexts impose significant pressure on network bandwidth, system interconnects, and memory capacity due to the enormous data volume~\cite{sun2021building, reis2020computing,gupta2021accelerating,glova20193d,de2021does}.

\paragraph{Functionality}
An HE system can reduce computation, memory, and storage overhead by encrypting data at the word level or limiting the depth of operations~\cite{sun2021building, mohassel2017secureml,zhang2024arcedb}. However, both approaches come at the cost of functionality. Word-level encryption lacks native bitwise operations, such as AND/OR, making it infeasible for fine-grained comparisons for database tasks. Without bit-level operations, database systems cannot directly compare ciphertexts, posing challenges for queries involving equality or range checks. Additionally, limiting the depth of operations means that complex computations will inevitably require bootstrapping to reset noise levels, ensuring the correctness of results. Without sufficient bootstrapping, deeper computations may become unreliable, restricting the types of queries the system can support.

\subsection{Alternatives}
\label{sec:alternatives}
Prior work has investigated other alternatives for making database systems secure or privacy-preserving. However, none achieve the same level as \LHEDB{}. 
\subsubsection{Trusted Hardware-Based Database}
\label{sec:tee}
Conventional secure database systems rely on encrypted data storage, secure key exchange mechanisms, and trusted execution environments (TEEs)~\cite{alwaysencrypted, enclavedb, ribeiro2018dbstore, bajaj2011trusteddb, vinayagamurthy2019stealthdb, 8946540, 9925569, sun2021building, wang2022operon}. These designs assume the TEE can temporarily store the decryption key and perform the necessary decryption on the encrypted data to compute the plaintext data. If any component in the TEE is compromised, the whole system becomes insecure. In addition, the attacker can learn the data access patterns and gain access to the data due to the limited space within the TEEs~\cite{eskandarian2017oblidb}. Oblivious RAM (ORAM)-based database~\cite{eskandarian2017oblidb, crooks2018obladi} can alleviate the data access pattern leakage problem but still relies on the trustworthiness of the underlying hardware. In contrast, HE does not need to trust any third-party hardware, and the entire system remains intact even though we offload computation to compromised hardware. 

\subsubsection{Hybrid Cryptosystem}
Hybrid cryptosystems integrate multiple cryptographic techniques, such as order-preserving encryption (OPE), deterministic encryption (DET), and partial homomorphic encryption (PHE)~\cite{popa2011cryptdb,shafagh2015talos}. While hybrid systems offer flexibility for different operations, they introduce vulnerabilities like information leakage and side-channel attacks, especially in DET and OPE schemes~\cite{naveed2015inference, zhang2024secure, oya2021hiding, cui2021privacy}. These vulnerabilities compromise the system's security in ways that HE avoids.

\subsubsection{Multi-Party Computation (MPC)}
MPC distributes a secret among parties, ensuring none possess the entire secret. Each party conducts computations on partial secrets, combined by the source for the desired outcome. Prior studies demonstrate MPC's feasibility in databases~\cite{he2015sdb, sepehri2015privacy, laud2018privacy, volgushev2019conclave, bater2016smcql, liagouris2021secrecy, burkhart2010sepia, wang2021secure}.

MPC and HE serve different use cases and are complementary rather than competing approaches: 

\textbf{Different Trust Models:} MPC requires trust assumptions about non-collusion between parties—if sufficient parties collude, security fails. HE relies solely on cryptographic hardness assumptions, requiring no trust in the computing party. 

\textbf{Different Operational Models:} MPC requires active participation and coordination among multiple parties throughout computation, with associated communication overhead. HE enables fully asynchronous computation by a single untrusted party, with no interaction after initial data encryption. 

\textbf{Different Regulatory and Architectural Fit:} Single-provider custody requirements, common in healthcare and finance, may make distributed MPC architectures legally or practically infeasible. Conversely, scenarios requiring joint computation over distributed private inputs naturally fit MPC. 

\textbf{Performance Trade-offs:} MPC achieves dramatically better computational performance—Secrecy~\cite{liagouris2021secrecy} reports 6 seconds for TPC-H Q6 on 8 million rows, while HE-based systems such as \BASELINE{} require 12,100 seconds for just 32K rows. This gap reflects fundamental architectural differences: MPC performs computations on secret shares with modest overhead, while FHE requires complex polynomial arithmetic on large ciphertexts.

We consider MPC-based databases as orthogonal to our work. Organizations choose between MPC and HE based on their specific trust model, regulatory constraints, and architectural requirements. When HE is necessary due to these constraints, our work demonstrates that careful system design can make HE-based databases practical, achieving orders-of-magnitude improvements over prior HE systems.

\subsubsection{Existing HE-enabled DBMS}
\label{sec:sub:he_approach}
Before \LHEDB{}, researchers have explored the potential of using HE in database workloads. However, none is ideal in addressing the performance, space/memory, and functionality challenges. 

Early proposals use somewhat Homomorphic Encryption (SHE) in retrieving encrypted records through conjunction queries~\cite{boneh2013private, ahmad2022pantheon, mughees2021onionpir, yi2012single, koirala2024summation}. However, as SHE only supports addition and multiplication, SHE-based systems can only achieve private information retrieval (PIR) but not general queries.

Recent projects~\cite{ren2022heda,he3db,zhang2024arcedb} strategically combine various HE schemes to enhance the versatility of HE-based systems. These works store data using bit-level encryption (TFHE) and apply logical operations for comparisons in \texttt{WHERE} clauses. For arithmetic operations, all of them dynamically transcipher encrypted data from TFHE to BFV for HEDA~\cite{ren2022heda} and to CKKS for HE$^3$DB~\cite{he3db} and ArcEDB~\cite{zhang2024arcedb}. However, the transciphering process is as expensive as bootstrapping, as it involves re-encrypting data under a new HE scheme using the secret key of the new scheme~\cite{chen2021efficient, bae2023hermes, lu2021pegasus}. HEDA further reports that 90\% of the query evaluation time is spent on the Parameter-Lifting Bootstrapping procedure. Prior work has also explored encoding data into a vector of field elements (VFE) to leverage data-level parallelism, demonstrating the potential to manage comparison latency at the millisecond level~\cite{tan2020efficient}. However, these previous approaches do not encompass the broader domain of query evaluations addressed in this paper.


\section{\LHEDB{} Design Overview}
\label{sec:overview}

\begin{figure*}
    \centering
    \includegraphics[width=\textwidth]{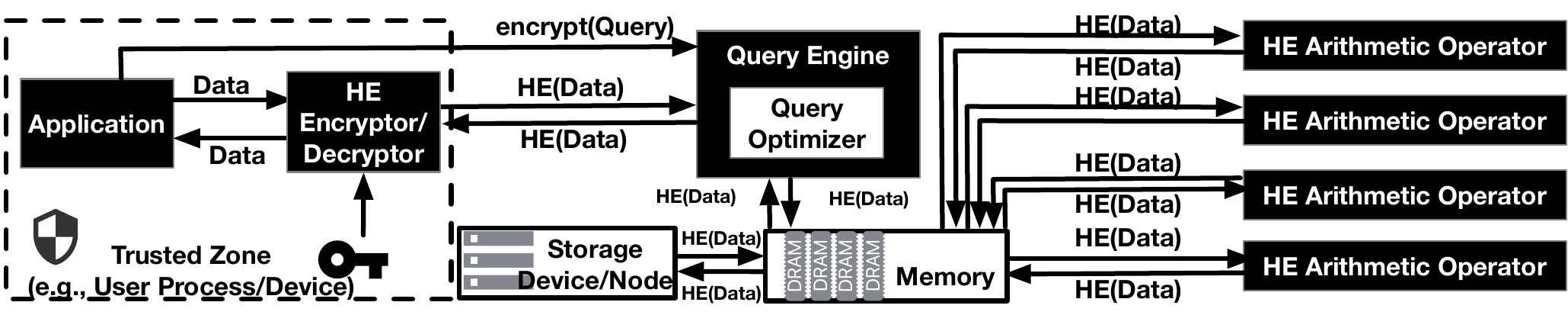}
    
    \vspace*{-0.1in}
    \caption{The system overview of \LHEDB{}}
    \label{fig:overview}
    \vspace*{-0.2in}
\end{figure*}

Figure~\ref{fig:overview} provides an overview of \LHEDB{}. It assumes the user application runs in a trusted zone (e.g., a private device or virtual machine) where only trusted software components can access the secret key. The application does not trust the system's interconnect, storage, or other resources. When \LHEDB{} receives a query to load or store data, it encrypts or decrypts batch ciphertext using the BFV scheme and the secret key. The query engine then translates operators into arithmetic operations supported by \LHEDB{}'s encoding. \LHEDB{} evaluates the noise budget based on the computational chain and seeks opportunities to minimize multiplicative depth, effectively shortening the chain. By reducing depth where possible, \LHEDB{} postpones the need for bootstrapping, improving overall execution efficiency. Throughout this process, \LHEDB{} does not require any key exchanges with any resource outside the trusted zone and maintains data in HE formats.

At a high level, the \LHEDB{} design adheres to three key principles to address the challenges of modern HE schemes in database systems:

\paragraph{A Constraint-Driven Scan-First Architecture: Defining a New Optimization Objective.}
\LHEDB{} stores data homomorphically encrypted using word-level HE batch encoding, placing multiple data points into a single ciphertext to avoid 10,000$\times$ data size expansion in bit-level. For instance, with an encryption parameter setting of a polynomial degree of 32,768, \LHEDB{} stores 32,768 data points in a single ciphertext instead of using 32,768 separate ciphertexts while enabling massive parallelism. However, simply applying word-level encoding creates a fundamental challenge: \textbf{no direct slot access}. The HE scheme operates in a SIMD-style, applying operations to all slots simultaneously. Isolating slot $i$ requires a multi-step process involving creating a mask vector, multiplying the entire ciphertext by this mask, and using rotations to move the result.

This constraint fundamentally breaks traditional database assumptions and imposes a \textbf{new architectural objective}: Positional access through masking costs the same as full scans or leaks access patterns, demonstrating a shift in complexity from logarithmic to linear.

\begin{itemize}[leftmargin=*,topsep=0pt,itemsep=2pt]
    \item \textbf{Index structures become ineffective:} While access to position $i$ is technically possible through masking, it requires processing the values in the ciphertext---incurring the same cost as scanning all elements. B-trees and hash tables thus lose their efficiency advantage when \texttt{array[i]} costs $O(n)$ instead of $O(1)$.
    
    \item \textbf{Data manipulation degrades to full rewrites:} Operations, such as swapping two elements or moving records between buckets, cannot be done in-place. Each "point update" requires creating masks, processing entire ciphertexts, and reconstructing the result---essentially rewriting the entire packed vector.
    
    \item \textbf{Security amplifies inefficiency:} Even if masking were cheap, using it to access specific positions would leak access patterns. To maintain security, we must process all positions anyway and mask the results.
\end{itemize}

Rather than attempting to preserve traditional access methods under this constraint, \LHEDB{} takes a different approach. Recognizing that conventional indexing provides no performance benefit when all operations require full ciphertext processing, we design a \textbf{scan-first architecture} that exploits the inherent parallelism of word-level encoding---processing 32,768 values per operation. This represents a specialized query processing architecture designed specifically for the unique characteristics of HE.

The implications of word-level HE constraints extend beyond simple data access to affect all traditional database optimizations. Materialized views, while technically feasible, suffer from the batching-versus-incremental tension. Consider a view with predicates (e.g., \texttt{SUM(price) WHERE price > 1000}). Incremental maintenance requires evaluating predicates on single rows, paying the full comparison cost without data-level parallelism from batching---the same computational cost for a single entry as for 32,768 entries. Materialized join results face prohibitive memory costs: a 1K$\times$1K join produces 7.4GB of ciphertexts. Materialized views only benefit narrow cases: (1) predicate-free aggregations requiring single ciphertext additions, (2) complex views where full recomputation cost exceeds incremental update overhead by orders of magnitude, or (3) storing aggregated join results rather than full joins.

Table~\ref{tab:plaintext-vs-fhe} quantifies how packing transforms database operations:

\begin{table}[h]
\centering
\small
\begin{tabular}{lcc}
\toprule
\textbf{Operation} & \textbf{Plaintext} & \textbf{FHE with Packing} \\
\midrule
Point lookup & $O(1)$ or $O(\log n)$ & $O(S)$ (process all slots) \\
Range scan & $O(\log n + k)$ & $O(n)$ (scan + mask) \\
Hash join & $O(n + m)$ & $O(n \cdot \lceil m/S \rceil)$ \\
Optimization target & I/O, CPU & Multiplicative depth \\
\bottomrule
\end{tabular}
\caption{Packing forces scan-based behavior in all FHE operations}
\label{tab:plaintext-vs-fhe}
\end{table}

These fundamental shifts in complexity---from logarithmic to linear, from I/O-bound to depth-bound---demonstrate that FHE requires not adaptation but complete rearchitecting of database query processing.

\paragraph{Translating logical operations into arithmetic operations to avoid transciphering overhead}
\LHEDB{} translates unsupported logical operations into arithmetic equivalents to avoid the need for transciphering between different HE schemes. Traditional HE-based systems often rely on transciphering between word-level and bit-level encryption to support logical operations such as AND, OR, and XOR. However, this process incurs significant latency and efficiency losses.

Specifically, transciphering introduces three major issues:
\begin{enumerate}
\item \textbf{Transciphering Latency:} Switching between word-level (e.g., BFV/CKKS) and bit-level (e.g., TFHE) encryption is costly, often involving re-encryption and bootstrapping—one of the most expensive HE operations. For instance, transciphering from LWE (TFHE format) to RLWE (BFV/CKKS format) is up to 100,000\x{} more expensive than a single homomorphic arithmetic operation~\cite{lu2021pegasus, bae2023hermes}, significantly slowing down query execution.
    \item \reviewed{\textbf{Loss of Batch Encoding:} Word-level encryption (e.g., BFV/CKKS) supports batching (e.g., processing 32,768 elements in one ciphertext), enabling parallelism. Transciphering to bit-level encryption removes this capability, leading to significantly slower performance.}
    \item \textbf{Frequent Bootstrapping:} Bit-level encryption schemes lack the "levels" present in word-level schemes, necessitating frequent bootstrapping to control noise accumulation. This adds additional overhead and further increases the latency of bit-level operations.
\end{enumerate}

To overcome these drawbacks, \LHEDB{} implements logical operations directly as arithmetic operations, eliminating the need for transciphering. While this increases the number of arithmetic operations, it retains the advantages of batch encoding, making the system far more efficient. Section~\ref{sec:HE_operators} provides a detailed explanation of how \LHEDB{} translates essential logical operations in database workloads into arithmetic operations.

\paragraph{Managing Noise to Reduce Bootstrapping.} 
Rather than relying on bootstrapping to control noise, \LHEDB{} uses LHE to manage the noise budget during query execution. Since bootstrapping can be 5,000× to 140,000× slower than standard homomorphic operations~\cite{al2023demystifying,OpenFHE,cheon2024dacapo,van2023fpt}, \LHEDB{} fundamentally reimagines query optimization for the FHE domain. Under leveled HE with packing, all operations are inherently scan-based---the optimizer cannot choose between access methods but must instead minimize multiplicative depth while scanning. This leads to counterintuitive design decisions: \LHEDB{} inverts traditional heuristics through Predicate Pull-Up and Mask-Injection Tuning, deliberately delaying filter applications when it reduces depth. By decomposing complex predicates into independent subgraphs and carefully choosing where to inject masks in join trees,  \LHEDB{} optimizes query plans to reduce multiplicative depth, thereby minimizing the frequency of bootstrapping. Section~\ref{sec:noise} provides further details on these noise-sensitive query optimizations.

\paragraph{Security Analysis}
\LHEDB{} assumes a semi-honest adversary who correctly executes protocols but may attempt to infer information from encrypted data—the standard model for encrypted databases~\cite{he3db,zhang2024arcedb,popa2011cryptdb,arasu2015transaction}. \LHEDB{}'s security relies on BFV's IND-CPA (Indistinguishability under Chosen Plaintext Attack) guarantee, ensuring ciphertexts reveal nothing about plaintexts, and RLWE (Ring Learning With Errors), which makes decryption computationally infeasible without the secret key.

\LHEDB{} exposes the following leakage, denoted as $L$:
\begin{itemize}[leftmargin=*,topsep=0pt,itemsep=2pt]
    \item \textbf{Schema information:} Table structure, column names, foreign key constraints
    \item \textbf{Static metadata:} Base table sizes, dictionary sizes for encoded columns  
    \item \textbf{Query structure:} Operator DAG shape (but not predicate values or selectivities)
    \item \textbf{Output structure:} Fixed $n$-slot ciphertext vectors for all operators
\end{itemize}

Operator-specific leakage includes:
\begin{itemize}[leftmargin=*,topsep=0pt,itemsep=2pt]
    \item \textbf{SELECT:} Masks non-matches with encrypted zeros; ciphertext structure remains unchanged, concealing selectivity. Zeros are dropped on the client-side.
    \item \textbf{GROUP BY:} Groups only on foreign key columns and always produces one aggregate per parent key (|P| ciphertexts, where |P| is already leaked via schema).
    \item \textbf{JOIN:} Behaves identically to GROUP BY—outputs a fixed number of ciphertexts based on key cardinality. Equality masks are computed in bulk, revealing only output count (already leaked via FK constraints), not which specific key pairs matched.
    \item \textbf{ORDER BY:} Scans the entire dictionary with fixed iterations, concealing value-frequency patterns.
    \item \textbf{Aggregations:} Return fixed-size results.
\end{itemize}

All operations utilize IND-CPA secure BFV ciphertexts, ensuring no information leaks beyond $L$. This leakage profile aligns with prior encrypted databases~\cite{popa2011cryptdb,poddar2016arx} while uniquely avoiding the need for trusted hardware and the overhead of scheme switching.
\section{\LHEDB{}}
\label{sec:architecture}

\LHEDB{} enhances the efficiency of HE for database workloads by leveraging the LHE BFV scheme, which offers benefits such as space efficiency, avoidance of bootstrapping, and batch encoding. However, simply applying the LHE BFV scheme is insufficient for DBMSs, as the LHE BFV scheme lacks the critical support of operators necessary for general-purpose database queries. This paper makes two significant contributions to the realization of LHE database systems. First, \LHEDB{} presents an innovative approach that efficiently supports the missing comparison operations, which are critical for database queries but are not natively supported by the BFV scheme. Second, \LHEDB{} introduces a novel noise-sensitive query optimization technique that manages noise accumulation during query execution while minimizing the invocation of bootstrapping. This section will describe both of these innovations in detail.

\subsection{Data Storage and Encoding}
\label{sec:data_encoding}
\LHEDB{} encrypts data using the BFV scheme with batch encoding, packing 32,768 values into a single ciphertext. This significantly improves storage efficiency compared to encrypting values individually—storing one ciphertext per batch instead of 32,768 separate ciphertexts. For example, fully packing and encrypting 32,768 entries results in 27 times the raw data size—0.27 MB of raw data expands to a 7.4 MB ciphertext.

\ignore{
\hl{While BFV supports arithmetic operations on packed ciphertexts (Section~\ref{sec:bfv_ops}) and standard data type encodings (Section~\ref{sec:data_type_encoding}), batch encoding introduces a unique challenge: handling partially-filled ciphertexts.}

{\subsubsection{Handling Padding with Valid Bits}
\label{sec:valid_bit}
\LHEDB{} introduces a \texttt{valid bit} column to differentiate real data from padded rows. When batch encoding, \LHEDB{} pads data if the number of entries falls short of the maximum vector size. However, padding with zeros can cause false positives in Boolean queries (e.g., mistakenly matching False values), while padding with random values can mislead queries into treating them as legitimate data.

To prevent such errors, \LHEDB{} multiplies query results by the valid bit column during filtering and aggregation. This approach ensures that queries process only real data while ignoring padded rows. By distinguishing valid entries from padding, the valid bit preserves query accuracy without adding significant computational overhead.
}}

\subsection{Supporting Database Queries}
\label{sec:HE_operators}
Unlike existing alternatives~\cite{ren2022heda, he3db, zhang2024arcedb} that rely on transciphering between word-level and bit-level encryption schemes, \LHEDB{} builds database operators entirely within BFV's arithmetic framework. By leveraging the comparison operations described in Section~\ref{sec:equality}, \LHEDB{} avoids the costly scheme conversions that dominate execution time in prior systems. This section describes how \LHEDB{} combines these arithmetic building blocks to construct efficient database query operators.

\subsubsection{Operator Support Overview}
Table~\ref{tab:operator-support} summarizes \LHEDB{}'s operator support. All operations work within BFV's word-level encryption, avoiding transciphering overhead.
\begin{table}
\centering
\caption{NSHEDB Operator Support and Complexity}
\label{tab:operator-support}
\small
\begin{tabular}{lcc}
\toprule
\textbf{Operator} & \textbf{Support} & \textbf{Complexity}\\
\midrule
Selection (WHERE) & \checkmark & $O(n/S)$ ops \\
Projection & \checkmark & $O(1)$ \\
COUNT/SUM & \checkmark & $\log S$ depth \\
AVG & Partial$^\dagger$ & returns (SUM, COUNT) \\
GROUP BY & \checkmark & $O(|G|\cdot n/S)$ ops \\
JOIN & \checkmark & $O(N\cdot \lceil M/S\rceil)$ ops \\
ORDER BY & \checkmark & $O(|D|\cdot n/S)$ ops \\
Boolean AND/OR/NOT & \checkmark & via arithmetic$^\ddagger$ \\
B-tree/Hash index & $\times$ & degrades to scan \\
Native division & $\times$ & use workarounds \\
\bottomrule
\end{tabular}\\[2pt]
\footnotesize\raggedright
$^\dagger$ Client-side division. 
$^\ddagger$ AND=$ab$, OR=$a{+}b{-}ab$, NOT=$1{-}a$.
\end{table}
These complexities reflect packing constraints: \textbf{JOIN} reduces $N{\times}M$ comparisons to $N{\cdot}\lceil M/S\rceil$ ciphertext operations (e.g., $1K{\times}1K$ join = 1000 ops vs $10^6$ scalar). \textbf{GROUP BY} requires $|G|$ equality checks processing $S$ values in parallel—expensive for high-cardinality columns. \textbf{ORDER BY} enumerates all $|D|$ domain values—practical only for small categorical domains. Each equality needs $\approx$16 multiplications (Section~\ref{sec:equality}), high but below bootstrapping cost.

\subsubsection{Supporting Database Query Operators}
\label{sec:query_operators}
\ignore{
\begin{figure*}
    \centering
    \includegraphics[width=\textwidth]{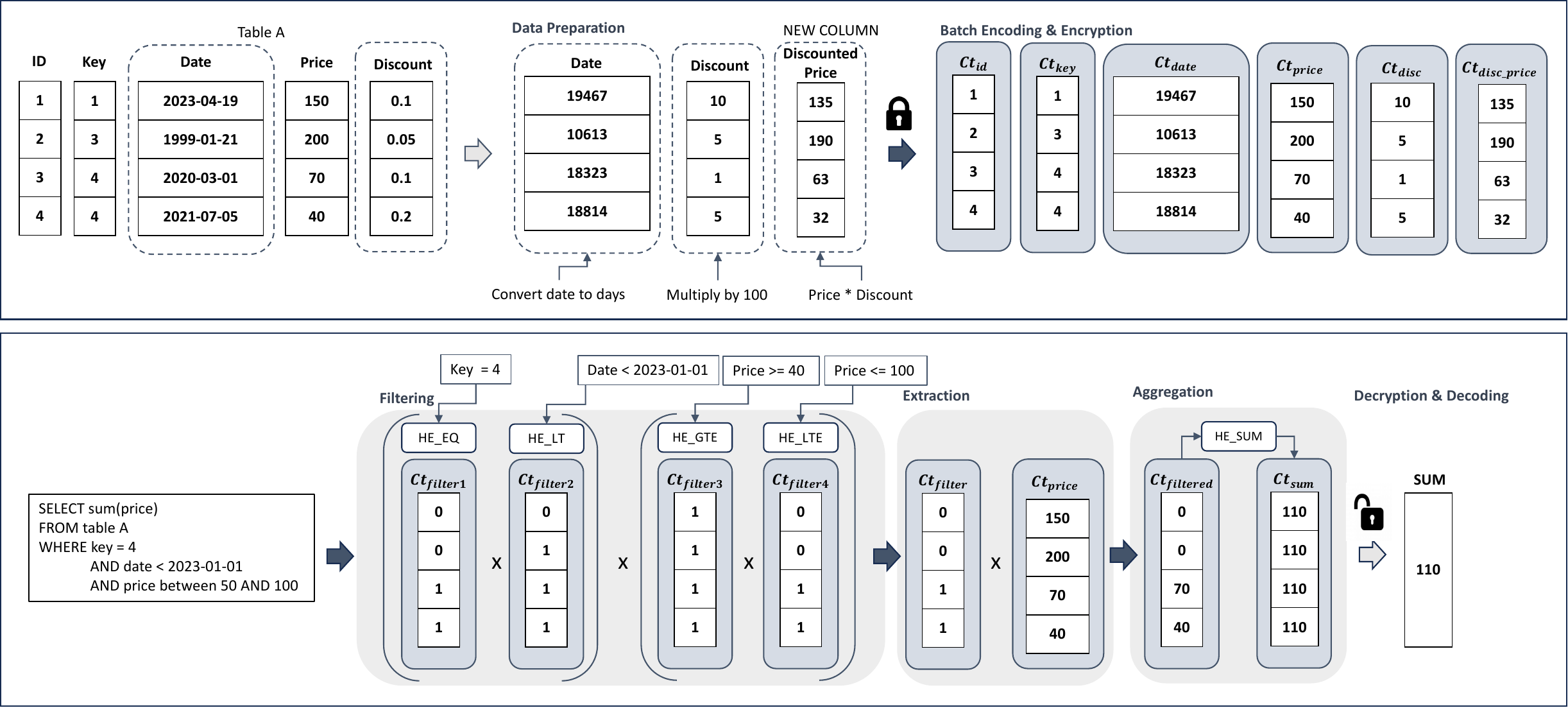}
    \caption{An illustration of the \LHEDB{} workflow for SELECT-and-AGGREGATE query operators.}
    \label{fig:exampleWorkFlow}
\end{figure*}

\begin{table}[t]
  \caption{Example Table A: ID is the primary key and Key is a foreign key referencing B.Key}
  \label{tab:a}
  \begin{tabular}{cccccc} 
    \toprule
    ID & Key & Date & Price & Discount & Shipping\\
    \midrule
    1 & 1 & 2023-04-19 & 150 & 0.1 & Standard\\
    2 & 3 & 1999-01-21 & 200 & 0.05 & Standard\\
    3 & 4 & 2020-03-01 & 70 & 0.1 & Express\\
    4 & 4 & 2021-07-05 & 40 & 0.2 & Standard\\
  \bottomrule
\end{tabular}
\end{table}

\begin{table}[t]
  \caption{Example Table B: Key is the primary key.}
  \label{tab:b}
  \begin{tabular}{ccc}
    \toprule
    Key & Size & Brand \\
    \midrule
    1 & 4 & 20\\
    2 & 6 & 1\\
    3 & 6 & 13\\
    4 & 2 & 7\\
  \bottomrule
  \end{tabular}
\end{table}
}
\ignore{
\begin{figure}[h]
\includegraphics[width=\columnwidth]{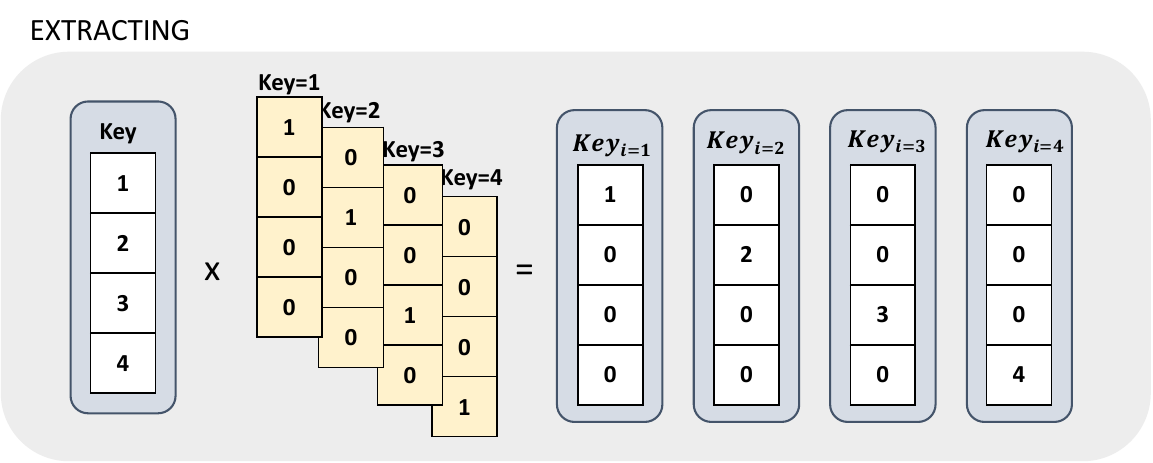}
    \caption{Extracting an entry within a ciphertext.}
    \label{fig:extract}
\end{figure}
}

\begin{figure*}[t]
  \includegraphics[width=\textwidth]{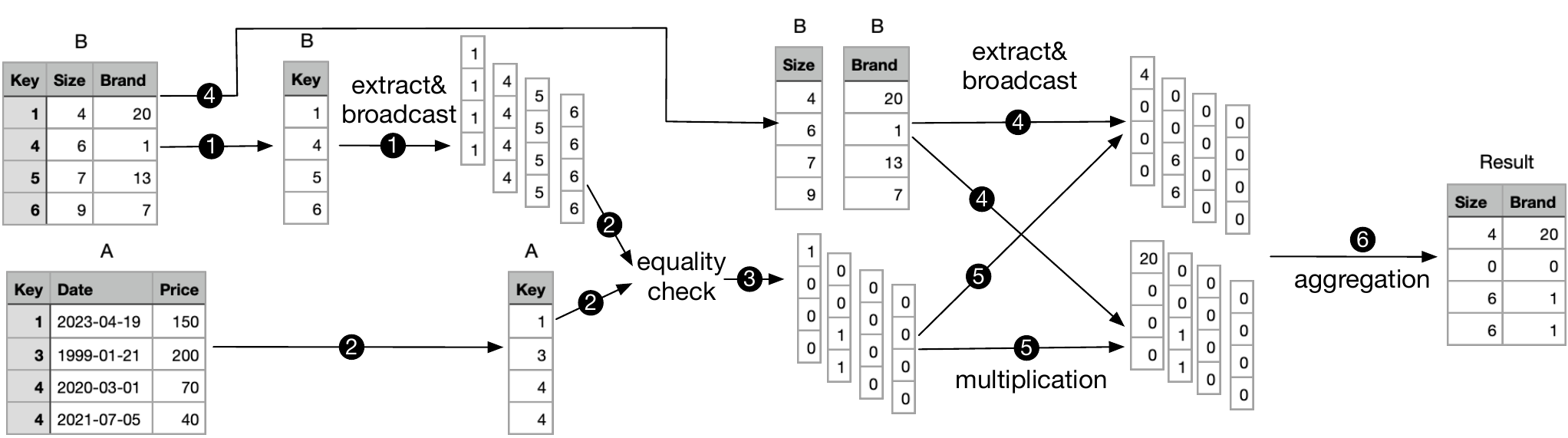}
  \caption{The process of \texttt{JOIN} in \LHEDB{}}
  \label{fig:groupby}
\end{figure*}

In this section, we describe how \LHEDB{} supports major SQL / relational algebra operators using HE. 
\ignore{
For a better explanation, we use example tables, Table \ref{tab:a} and Table \ref{tab:b}, to illustrate example queries or explain the mechanisms of operators.}

\paragraph{\texttt{SELECT} operator}

\LHEDB{} supports the \texttt{SELECT} operator ($\sigma$) by leveraging its unique comparison algorithms across HE-encrypted columns of interests. Since the third-party computing resource cannot decrypt the numerical results of HE computations to evaluate equality or inequality conditions, \LHEDB{} multiplies the encrypted columns with the results of equality or inequality evaluations under the \texttt{SELECT} condition.

For example, consider the following query:

\begin{lstlisting}[
          language=SQL,
          showspaces=false,
          basicstyle=\ttfamily\small,
          columns=fullflexible,
          frame=single,
          breaklines=true,
          numbers=left,
          numberstyle=\tiny,
          commentstyle=\color{gray}
        ]
SELECT A.price FROM A WHERE A.ID == A.key;
\end{lstlisting}

\LHEDB{} performs the following HE computation on the elements corresponding to the same record in the HE-encrypted columns:

\begin{equation} 
\label{eq:select}
    A.price \times EQ(A.ID, A.key)
\end{equation}

where $EQ$ denotes the equality function from Section~\ref{sec:equality}. For records where $A.ID == A.key$, the comparison function $EQ(A.ID, A.key)$ returns a ciphertext with encrypted "1"s in the corresponding positions. For records that do not match the condition, $EQ(A.ID, A.key)$ returns encrypted "0"s, effectively zeroing out those elements in the ciphertext. By multiplying the resulting ciphertext from the equality evaluation with the HE-encrypted $A.price$, only the exact values of $A.price$ that meet the condition are retained in the final ciphertext for the \texttt{SELECT} query.

\paragraph{Aggregation: COUNT, SUM, and AVG}

\LHEDB{} also supports aggregations over the results of a \texttt{SELECT} operator in the following ways.

\textbf{COUNT.}
Supporting \texttt{COUNT} requires overcoming the inability to directly access individual elements within batch-encoded ciphertexts. \LHEDB{} uses repeated HE rotation operations with a doubling pattern: rotate by 1 and add, then by 2 and add, then 4, 8, and so on.

For example, with 8 slots $[v_0, \ldots, v_7]$:
\begin{itemize}
\item Step 1: Rotate by 4, add $\Rightarrow [v_0{+}v_4, v_1{+}v_5, v_2{+}v_6, v_3{+}v_7, \ldots]$
\item Step 2: Rotate by 2, add $\Rightarrow [v_0{+}v_4{+}v_2{+}v_6, v_1{+}v_5{+}v_3{+}v_7, \ldots]$
\item Step 3: Rotate by 1, add $\Rightarrow [\sum_i v_i, \sum_i v_i, \ldots, \sum_i v_i]$
\end{itemize}

This rotate-sum reduction requires only $\log n$ steps because each step doubles the values accumulated. For \texttt{COUNT}, this applies to comparison result ciphertexts containing encrypted 1s and 0s. The final ciphertext contains the exact count in every slot.

\smallskip
\noindent
\textbf{SUM. }
\LHEDB{} applies a similar approach for \texttt{SUM} as it does for \texttt{COUNT}. However, instead of accumulating comparison results, \LHEDB{} multiplies the encrypted column of interest with the comparison result ciphertext and accumulates the actual values of the records using the same doubling rotation pattern.

\smallskip
\noindent
\textbf{AVG.}
Since BFV/BGV schemes lack division operations, \LHEDB{} employs three workarounds for \texttt{AVG} queries:
(1) \textit{Query Rewriting:} Transform predicates to avoid division—e.g., \texttt{salary > AVG(salary)} becomes \texttt{salary $\times$ COUNT(salary) > SUM(salary)}.
(2) \textit{Propagating Aggregates:} Pass \texttt{(SUM, COUNT)} pairs through the query plan instead of computing averages in intermediate results.
(3) \textit{Final Computation:} Perform actual division client-side after decryption using the \texttt{SUM} and \texttt{COUNT} components.

\paragraph{Range and set queries: \texttt{IN} and \texttt{BETWEEN}}
\noindent

\noindent
\textbf{IN (set membership). }
\LHEDB{} evaluates the \texttt{IN} operation by applying equality functions on the given set of numbers and adds all equality results. If $\mathbb{S}$ represents the given set of numbers, \LHEDB{} implements the following mathematical function to fulfill the evaluation of whether $x$ is in $\mathbb{S}$. 
\begin{equation}
\begin{aligned}
\label{eq:in}
IN(x, \mathbb{S})& = &\sum_{y \in \mathbb{S}} EQ(x,y) 
\end{aligned}
\end{equation}
Despite the execution time being linear to the number of elements in $\mathbb{S}$, the latency is still significantly lower than that of mechanisms relying on transciphering.

\smallskip
\noindent
\textbf{BETWEEN. }
\LHEDB{} computes the \texttt{BETWEEN} operation as the product of two comparison results, deriving $\leq$ and $\geq$ outcomes by combining the comparison functions from Section~\ref{sec:equality}.

\paragraph{\texttt{JOIN} and \texttt{GROUP BY}}

\LHEDB{} implements the \texttt{JOIN} (i.e., equi-joins $\bowtie$) and \texttt{GROUP BY} operators using equality comparisons across the entries in the targeted attribute(s).

\smallskip
\noindent
\textbf{GROUP BY. }
\LHEDB{} implements the \texttt{GROUP BY} operation by performing equality checks on each value in the group. First, \LHEDB{} retrieves metadata to determine the number of distinct values, $n$, in the grouping column. Since all values are already encoded into a contiguous integer range from $1$ to $n$, \LHEDB{} directly operates on these unique IDs, ensuring consistent processing.

\LHEDB{} then performs equality checks for each distinct value in the target column, resulting in $n$ ciphertexts, each containing encrypted “1''s for matching elements and encrypted “0''s for non-matching elements. For multi-column \texttt{GROUP BY}, \LHEDB{} generates permutations of the grouped values. For example, with columns $|G1| = 2$ and $|G2| = 3$, sorting by \texttt{GROUP BY} G1, G2 yields permutations like {(1,1), (1,2), (1,3), (2,1), (2,2), (2,3)}. Reversing the order to \texttt{GROUP BY} G2, G1 results in different permutations.

\LHEDB{} then performs equality checks on the columns for each permutation and multiplies the results to identify elements that meet the specified conditions.

\smallskip
 \noindent
 \textbf{JOIN. }
\LHEDB{} supports (i) two-way equi-joins, (ii) $k$-way \emph{star} joins ($k\!\ge 2$), and (iii) semi-/nested joins expressed with
\texttt{EXISTS}/\texttt{IN}. Every variant is built from three HE steps: (1) \texttt{Extract+Broadcast} the reference-table key; (2) homomorphic equality to mask the fact table; and (3) \texttt{Multiply} the mask with projected attributes followed by \texttt{Add} across rows.

As an illustrative example, consider the simplified SQL query demonstrating a basic JOIN scenario:
\begin{lstlisting}[
 language=SQL,
 showspaces=false,
 basicstyle=\ttfamily\small,
 columns=fullflexible,
 frame=single,
 breaklines=true,
 numbers=left,
 numberstyle=\tiny,
 commentstyle=\color{gray}
 ]
 SELECT size, brand FROM A, B WHERE A.key = B.key;
 \end{lstlisting}
Figure~\ref{fig:groupby} illustrates the detailed process: \circled{1} \LHEDB{} applies the \texttt{Extract} and \texttt{Broadcast} operations on each element from \texttt{B.key}, and \circled{2} performs equality checks between \texttt{A.key} and each ciphertext created by \texttt{Broadcast} from \texttt{B.key}. \circled{3} This creates a set of HE ciphertexts indicating matching records in table \texttt{A}. Next, \circled{4} \LHEDB{} applies the \texttt{Extract} and \texttt{Broadcast} operations to the \texttt{size} and \texttt{brand} columns in table \texttt{B}, and \circled{5} multiplies these extracted attributes with the equality check results from step \circled{3}. This process produces ciphertexts containing the matched entries for each attribute. Finally, \circled{6} \LHEDB{} aggregates these ciphertexts to yield the final JOIN output.
Recognizing memory constraints, \LHEDB{} optimizes JOIN operations by merging intermediate JOIN and aggregation steps when possible, reducing intermediate storage requirements and improving computational efficiency. This optimization is particularly beneficial when handling more complex JOIN patterns such as multi-way equi-joins or nested JOIN structures.

Recognizing memory challenges in joining attributes from different tables, \LHEDB{} fuses the join and aggregation processes to minimize memory consumption wherever possible. This approach reduces data duplication and intermediate storage, streamlining operations and enhancing efficiency, particularly for aggregation-focused joins.

\subsubsection{ORDER BY}
\label{sec:sortby}
\LHEDB{} sorts encrypted data homomorphically by using equality checks on all possible values in the encrypted domain. Since direct comparisons are not feasible in homomorphic encryption, \LHEDB{} generates equality masks for each distinct value, marking matching records with encrypted "1"s and others with encrypted "0"s. For each value, \LHEDB{} applies this mask to extract and aggregate matching records sequentially, constructing the final sorted sequence. This process is repeated in increasing order for ascending sorts and in reverse order for descending sorts. For multi-column sorting, \LHEDB{} first sorts by the highest-priority column, then applies the same process to resolve ties based on the next column.

\subsection{Noise-Sensitive Query Optimization}
\label{sec:noise}
\begin{table}[t]
\centering
\small
\begin{tabular}{ll}
\toprule
\textbf{Operation} & \textbf{Multiplicative Depth} \\
\midrule
Equality Check & $\lceil \log(p-1) \rceil$ \\
Between/In & $\lceil \log(p-1) \rceil + \lceil \frac{\log(k)}{p} \rceil$ \\
Aggregation & $\frac{\log(n)}{p}$ \\
Join & $\lceil \log(p-1) \rceil + 1$ \\
Group by/Order by & $\lceil \log(p-1) \rceil$ \\
\bottomrule
\end{tabular}
\caption{Multiplicative Depth Analysis of Operations in \LHEDB{} Implementation.}
\label{table:noise_levels}
\end{table}

\LHEDB{} optimizes query execution by minimizing noise accumulation and reducing the need for costly bootstrapping. Since multiplications contribute to noise growth the most, while additions increase noise more gradually—by a factor of $\frac{1}{p}$, where $p$ represents the number of distinct values the system can encode—\LHEDB{} reduces multiplication depth by reordering operations and breaking down multiplication chains to limit noise growth at both the intra- and inter-operator levels.

\subsubsection{The Noise Escalation of Operators and Optimizations} 
\LHEDB{} optimizes each query operator by evaluating multiplications independently, avoiding chained dependencies on previous results. Table~\ref{table:noise_levels} summarizes the noise levels for each operator. Below, we detail the optimizations for several critical operators.

\smallskip
\noindent
\paragraph{Equality checks}
\LHEDB{} translates equality checks into arithmetic operations using Equation~\ref{eq:equality}. To minimize noise, \LHEDB{} reduces the number of required multiplications to $\lceil \log (p-1) \rceil$ by applying the exponentiation by squaring method~\cite{expSquaring}.

\smallskip
\noindent
\paragraph{BETWEEN/IN}
\LHEDB{} implements \texttt{BETWEEN/IN} by evaluating multiple equality checks independently, which limits noise escalation to $\lceil \log (p-1) \rceil$ multiplications. During aggregation, \LHEDB{} employs a divide-and-conquer algorithm to perform additions independently, avoiding the reliance on sequential results like in a traditional chain of operations. further reducing noise to $\lceil \frac{\log(k)}{p} \rceil$, where $k$ represents the size of the set for \texttt{IN} or the range for \texttt{BETWEEN}. This optimization limits the worst-case multiplication depth to $\lceil \log (p-1) \rceil + \lceil \frac{\log(k)}{p} \rceil$.

\smallskip
\noindent
\paragraph{AGGREGATION Operations}
\LHEDB{} implements aggregation operations, including \texttt{SUM} and \texttt{COUNT}, through rotations and additions. In our HE system, the noise level introduced by rotations is negligible~\cite{sealcrypto}. \LHEDB{} applies a divide-and-conquer algorithm to efficiently summarize vectors after different shift amounts. The noise level increases by $\frac{\log(n)}{p}$ as rotations cycle through the padding elements across all records within the column, where $n$ represents the maximum number of elements stored in a ciphertext.

\smallskip
\noindent
\paragraph{JOIN}
\LHEDB{} implements the \texttt{JOIN} operation in two steps. First, it identifies matching entries, producing a ciphertext of matching indexes. Then, it multiplies the joining column with this resulting ciphertext. The first stage uses equality checks, which escalate the noise level by $\lceil \log (p-1) \rceil$ multiplications. The second stage performs one HE multiplication. Therefore, in the worst case, the noise level escalation in \LHEDB{}'s \texttt{JOIN} operation equals $\lceil \log (p-1) \rceil + 1$ multiplications.

\smallskip
\noindent
\paragraph{GROUP BY/ORDER BY}
\LHEDB{} implements the \texttt{GROUP BY/ORDER BY} operation by performing equality checks on permutations of possible combinations and counting the matching records. \LHEDB{} performs equality checks, using $\lceil \log (p-1) \rceil$ multiplications to generate the resulting ciphertexts.

\subsubsection{Inter-Operator Optimizations}
\begin{figure}[t]
    \centering
    \begin{minipage}[b]{0.45\textwidth}
        \centering
        \includegraphics[width=\linewidth]{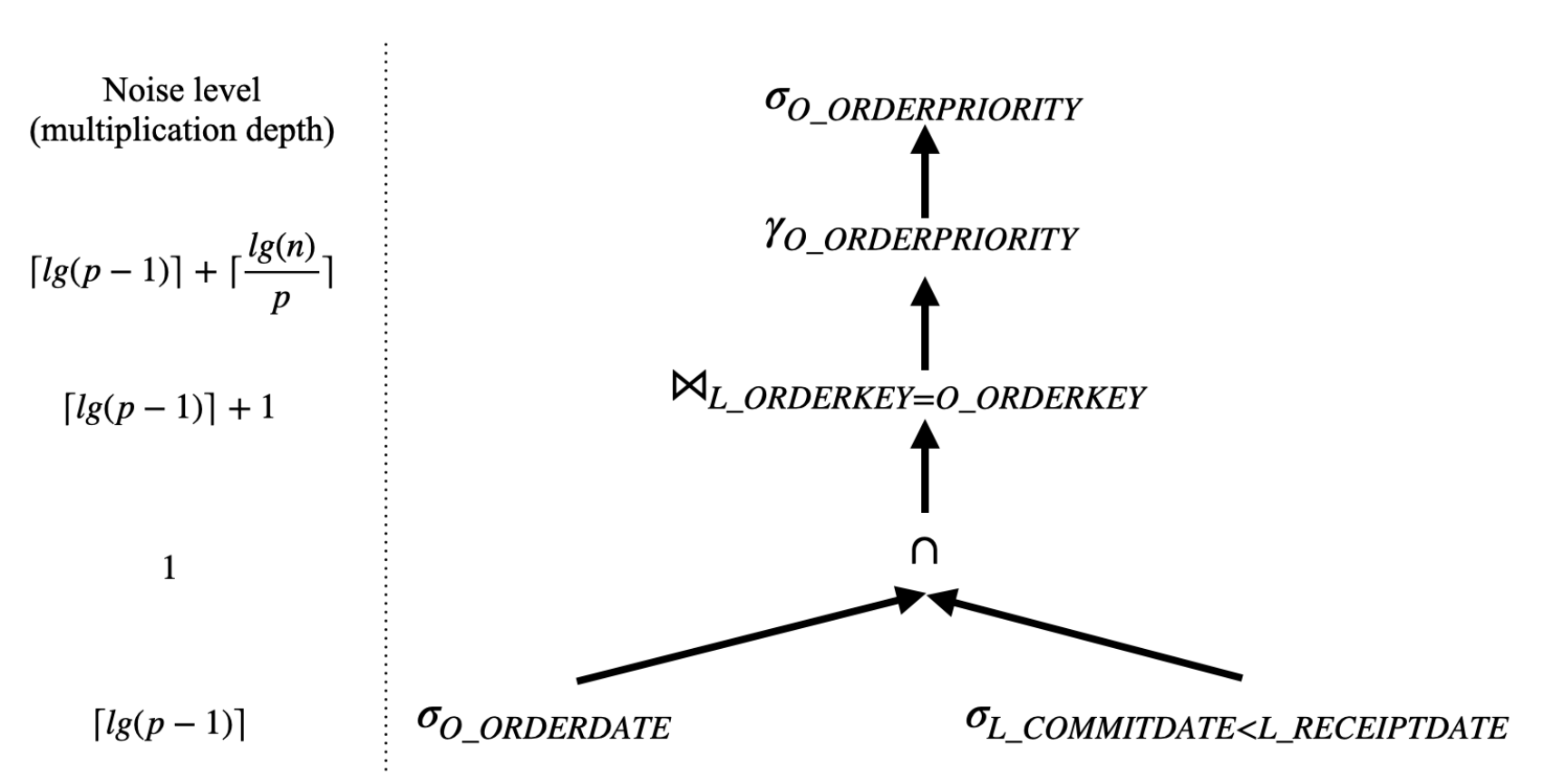}
        \par\vspace{2mm} 
        (a) Original Plan
    \end{minipage}
    \hfill 
    \begin{minipage}[b]{0.45\textwidth}
        \centering
        \includegraphics[width=\linewidth]{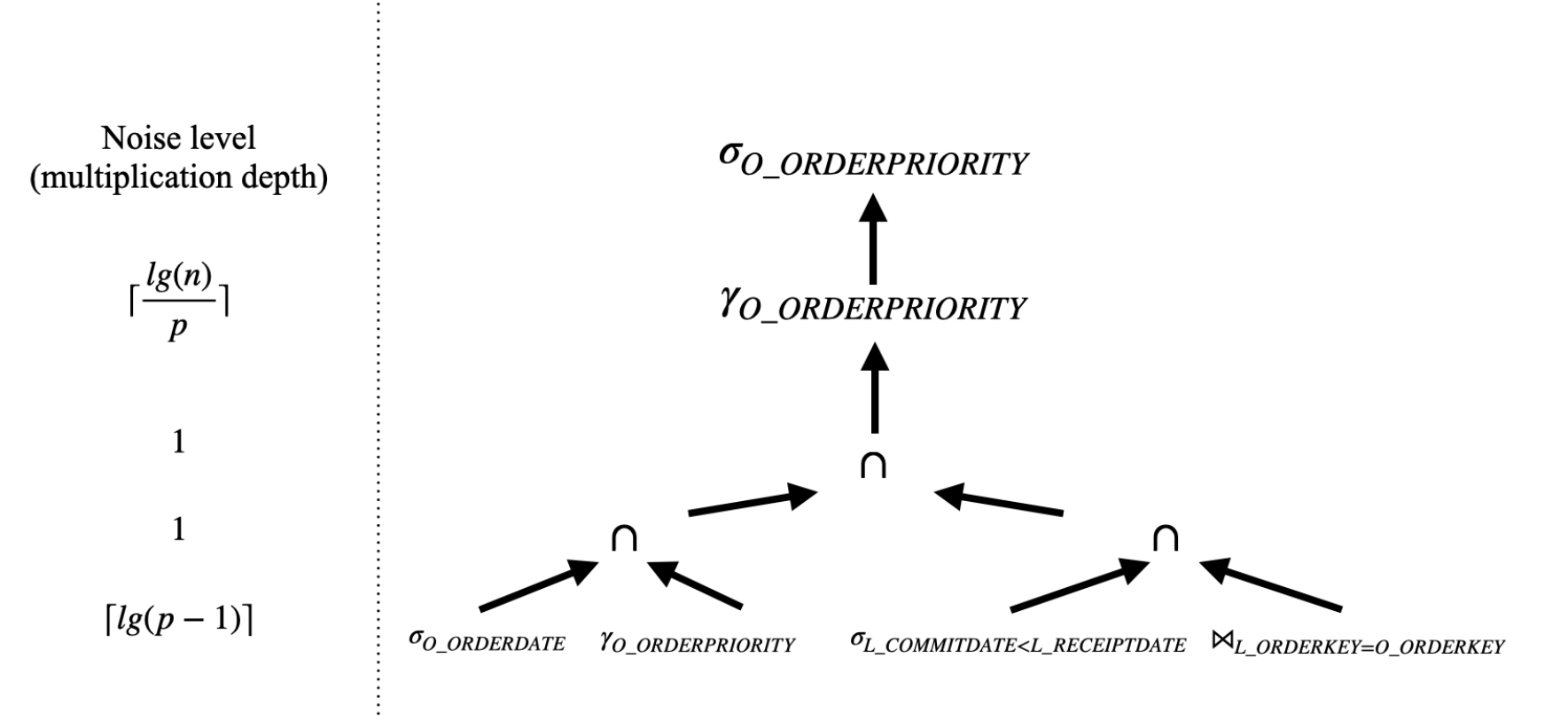}
        \par\vspace{2mm}
        (b) Noise-optimized Plan
    \end{minipage}
    
    \vspace{-0.1in}
    \caption{Noise optimization example. Plan~(b) executes the \texttt{SELECT} and \texttt{JOIN} operators independently and then intersects their outputs, saving roughly one equality check in multiplicative depth compared with the sequential execution in plan~(a).}
    \label{fig:inter-optimization}
\end{figure}

\LHEDB{}’s goal is to cap noise \emph{growth} by restructuring the query plan. Instead of executing operators in a deep, sequential pipeline, \LHEDB{} splits them into independent paths, thereby reducing multiplicative depth and avoiding unnecessary bootstraps.

\paragraph{Predicate Pull-Up \& Mask-Injection Tuning}

Classic predicate push-\emph{down} shrinks intermediate data early. Under homomorphic encryption, however, filtering leaves ciphertext sizes unchanged; irrelevant tuples are merely multiplied by \emph{predicate masks} and turned into encrypted zeros. The dominant cost therefore shifts from I/O to the \emph{multiplicative depth} that provokes noise growth and bootstrapping.
Given a mask, such as a date filter (\(\sigma_{1994\le O.\textit{date}<1995}\)) or a join equality (\(n\_regionkey=r\_regionkey\)), the planner must decide where to inject it: \emph{early}, pruning later work but lengthening the deepest multiplication chain, or \emph{late}, duplicating multiplications while shortening that chain.

To balance this trade-off, \LHEDB{} proposes two optimizations, \emph{Predicate Pull-Up} and \emph{Mask-Injection Tuning}, where predicate pull-up postpones mask multiplications and mask-injection tuning injects mask multiplications only at the latest depth that still satisfies the noise budget. Every filter or comparison runs in its own sub-graph to yield a mask ciphertext, and operators such as \texttt{GROUP BY} execute directly on the original columns without prior filtering. In an FK–PK \emph{chain}, the same mask can be injected at \emph{any} nesting depth—from the outermost loop down to the innermost loop. Injecting lower shortens the longest path (consuming less noise budget) but forces the mask to be extracted and broadcast at each additional level. Inside a single join tree, the planner faces the same depth-versus-work trade-off that predicate pull-up solves across subgraphs.

\paragraph{Cost-and-Decision Model.}
Consider a plan fragment whose longest path consists of \(m\) homogeneous stages (filters, joins, or a mix).
Let \(d_s\) denote the multiplicative depth \emph{per} stage (\(\approx\lceil\log(p-1)\rceil+1\) for an equi-join, \(1\) for a single multiplication, etc.). If the mask is injected after stage~\(i\) (\(0\le i\le m\)),\footnote{\(i=0\): inject immediately; \(i=m\): inject at the end.} then
\[
D_i=(m-i)d_s,
\qquad
\Delta\textit{MUL}_i=i\,C_{\text{mul}}.
\]
The incremental cost is
\[
\text{Cost}(i)=
(m-i)C_{\text{mul}}+\Delta\textit{MUL}_i
+\bigl[D_i>B_{\text{noise}}\bigr]\,C_{\text{boot}},
\]
with \(C_{\text{boot}}\gg C_{\text{mul}}\).
The planner chooses
\[
i^\star=
\begin{cases}
\max\{\,i\mid D_i\le B_{\text{noise}}\}, & \text{if feasible};\\
m, & \text{otherwise (pay one bootstrap).}
\end{cases}
\]

A single knob thus governs both \emph{predicate pull-up} (\(m=1\)) and
\emph{mask-injection tuning} (larger \(m\)).

\paragraph{Noise-Optimized Plan Construction}
A single depth-first traversal of the algebraic DAG applies three rewrite rules:
\begin{enumerate}[nosep,leftmargin=*]
  \item \textbf{R1 Mask isolation} — rewrite every
        $\sigma_p(R)$ as $\langle R, M_p\rangle$
        to compute the predicate once and store its mask.
  \item \textbf{R2 Independent evaluation} — when two masks touch disjoint
        attributes, split
        $\sigma_{p\land q}(R)$ into
        $\sigma_p(R)\;\cap\;\sigma_q(R)$, halving the longest chain.
  \item \textbf{R3 Late injection} — push each mask past joins until the
        deepest path length is $\le B_{\text{noise}}$
        (the cost model yields $i^{\star}$); inject once, then intersect or
        aggregate without further multiplications.
\end{enumerate}
Repeated R1 + R2 realize \emph{Predicate Pull-Up}; R3 at depth $i^{\star}$ realizes \emph{Mask-Injection Tuning}.
The whole pass runs in $O(|V|\!+\!|E|)$ time and short-circuits the longest multiplication chains while keeping extra multiplications cheap.

\paragraph{Example (TPC-H Q4).}
Figure~\ref{fig:inter-optimization} visualizes the impact of our noise optimization on TPC-H~Q4.  
In the baseline plan (Fig.~\ref{fig:inter-optimization}\,(a)), a sequential \texttt{JOIN}–\texttt{WHERE} pipeline incurs
\[
D_{\text{orig}} = 3\lceil\log(p-1)\rceil + \tfrac{\log n}{p} + 2 .
\]

After applying \emph{Predicate Pull-Up} and \emph{Mask-Injection Tuning} with \(i^\star = 1\) (Fig.~\ref{fig:inter-optimization}\,(b)), the date filter is injected only \emph{after} the join, reducing the depth to

\[
D_{\text{opt}} = \lceil\log(p-1)\rceil + \tfrac{\log n}{p} + 2 ,
\]

thereby eliminating two bootstraps whenever
\(D_{\text{opt}} \le B_{\text{noise}}\).

\paragraph{Take-away.}
\emph{Predicate pull-up} splits a plan into independent sub-graphs, while \emph{mask-injection tuning} reorders multiplications inside a join path.  Together, they respect the noise budget across hardware and parameter sets, choosing extra multiplications whenever they are cheaper than bootstrapping.

\section{Result}
\label{sec:result}

We implemented \LHEDB{} and evaluated it on a real system using TPC--H
queries. The prototype delivers up to a \textbf{659\x{} speed-up} over the
strongest prior HE database and consumes only \textbf{1.3 \%} of its
storage space. This section describes the evaluation platform,
workload, and results.

\subsection{Experimental Methodology}

\paragraph{Hardware and cryptographic parameters.}
All experiments run on a 16-core Intel Raptor-Lake CPU (5.1 GHz boost)
with 128 GB RAM.  Homomorphic operators use Microsoft
SEAL \cite{sealcrypto}.  Parameters follow the HE
Standard \cite{hestandard}: polynomial degree
$n{=}32\,768$, coefficient modulus $\log Q{=}881$, and plaintext modulus
$p{=}65\,537$, yielding 128-bit security.

\paragraph{Baselines.}
We compare \LHEDB{} with the hybrid-HE engines
\BASELINE{} \cite{he3db} and ArcEDB \cite{zhang2024arcedb}.

\paragraph{Dataset and query set.}
All eight TPC-H tables at scale factor 1 are loaded; the largest,
\texttt{LINEITEM}, is sampled to 32 K rows and related tables are scaled
proportionally. Note that this limited table size is primarily due to the space requirements of \BASELINE{} and ArcEDB, whereas \LHEDB{} can support larger workloads. We benchmark nine widely studied queries—Q1, 4, 5\footnote{We do not show Q5 in the later result figures because the theoretical runtimes of the two state-of-the-art baselines cannot finish this query within reasonable timeframe (more than a year) on our testbed. \LHEDB{} achieves 13,000\x{} speedup overall the projected runtimes of \BASELINE{} and ArcEDB. However, since the runtimes of the baselines are projected, we opt out to list Q5 in result figures for fairness.}, 6, 8, 12, 14, 17, 19—that together span the key access patterns in TPC-H:

\begin{itemize}[leftmargin=1em,itemsep=2pt]
  \item \textbf{Scan + aggregation} (Q1, Q6, Q14) – pipelines with SUM,
        AVG, COUNT, \texttt{GROUP BY}, and \texttt{ORDER BY}.
  \item \textbf{Inner and multi-way equi-joins} (Q5, Q8, Q12, Q14, Q19)
        – two-table stars up to a six-table join, all with predicate
        push-down.
  \item \textbf{Semi- / nested joins} (Q4, Q17) – patterns expressed
        with \texttt{EXISTS} and correlated subqueries.
  \item \textbf{Complex predicates} (Q19) – disjunctions,
        \texttt{BETWEEN}, and \texttt{IN} tests across joined tables.
\end{itemize}

The workload therefore exercises scans, filters, arithmetic aggregates,
semi-joins, three- to six-way joins, correlated subqueries, and
multi-predicate filters—providing a balanced view of \LHEDB{}’s
performance without redundant cases.

\subsection{Single-Threaded Performance and Noise-Sensitive Query Optimization}
\label{sec:performance-optimization}

\subsubsection{Primitive Operations Performance}

Table~\ref{tab:primitive-times} presents the absolute execution times for primitive database operations, demonstrating the efficiency of \LHEDB{} at the operator level. \LHEDB{} achieves sub-millisecond performance for COUNT, SUM, equality, and IN operations, while comparison operations require only 3.66ms. These improvements stem from avoiding bit-level transciphering and leveraging batch encoding.

\begin{table}[h]
\centering
\small
\begin{tabular}{lccc}
\toprule
\textbf{Operation} & \textbf{\BASELINE{} (ms)} & \textbf{ArcEDB (ms)} & \textbf{\LHEDB{} (ms)} \\
\midrule
COUNT & 1.27 & 1.27 & 0.04 \\
SUM & 1.27 & 1.27 & 0.04 \\
Equality (=) & 283.33 & 16.00 & 0.09 \\
Comparison (<, >) & 150.83 & 16.00 & 3.66 \\
Comparison ($\le, \ge$) & 143.07 & 16.00 & 3.66 \\
BETWEEN & 287.35 & 33.69 & 7.32 \\
IN & 283.33 & 16.00 & 0.09 \\
GROUP BY & 283.33 & 16.00 & 0.09 \\
\bottomrule
\end{tabular}
\caption{Execution time of primitive database operations}
\label{tab:primitive-times}
\end{table}

\begin{figure}[t]
    \centering
    \begin{minipage}[b]{0.48\textwidth}
        \centering
        \includegraphics[width=\linewidth]{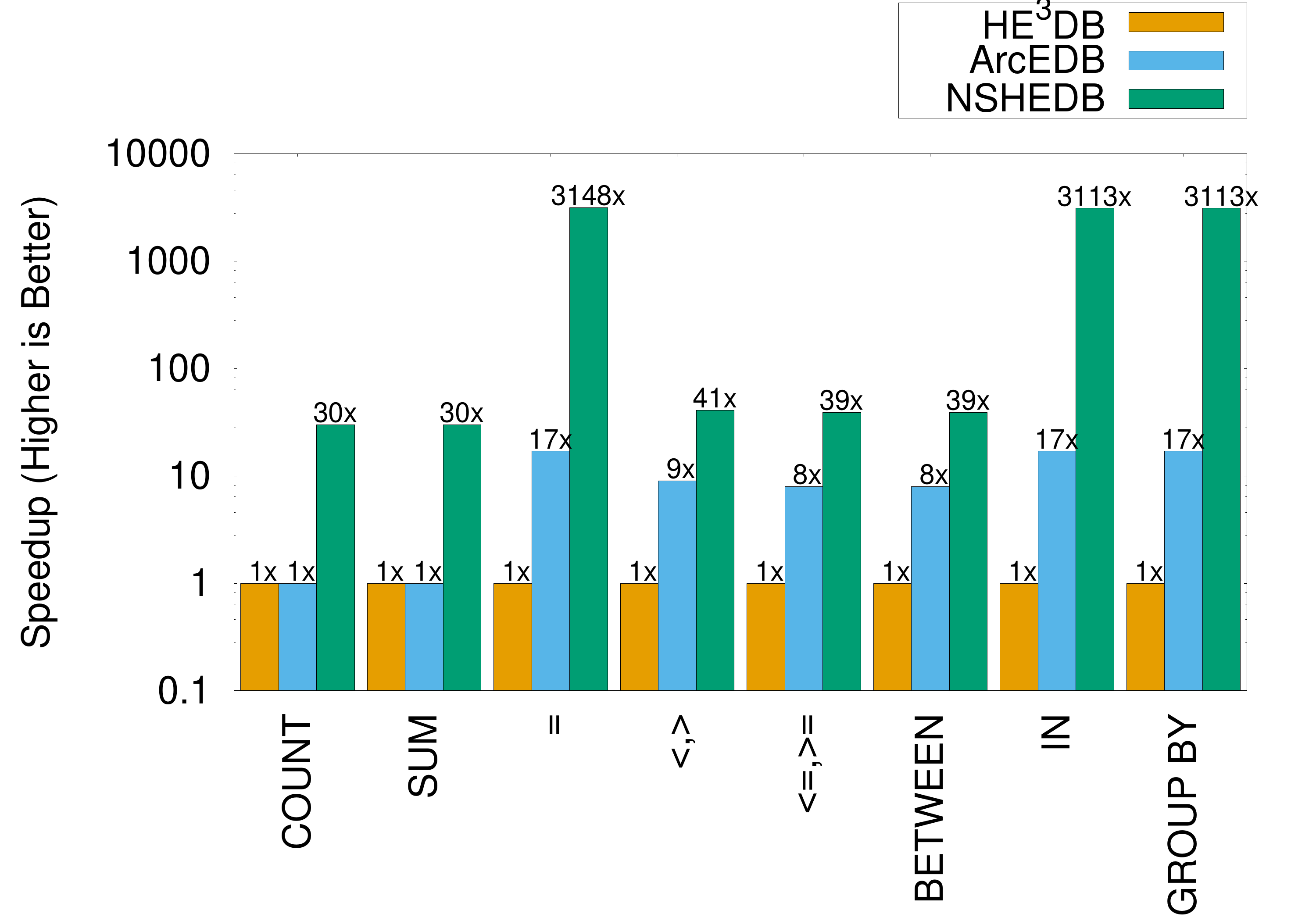}
        \vspace*{-0.2in}
        \caption{Speedup for basic database operations on \LHEDB{} compared to baselines. \LHEDB{} achieves speedups up to 3,148\x{} over \BASELINE{}.}
        \label{fig:microbenchmark}
    \end{minipage}
    \hfill 
    \begin{minipage}[b]{0.48\textwidth}
        \centering
        \includegraphics[width=\linewidth]{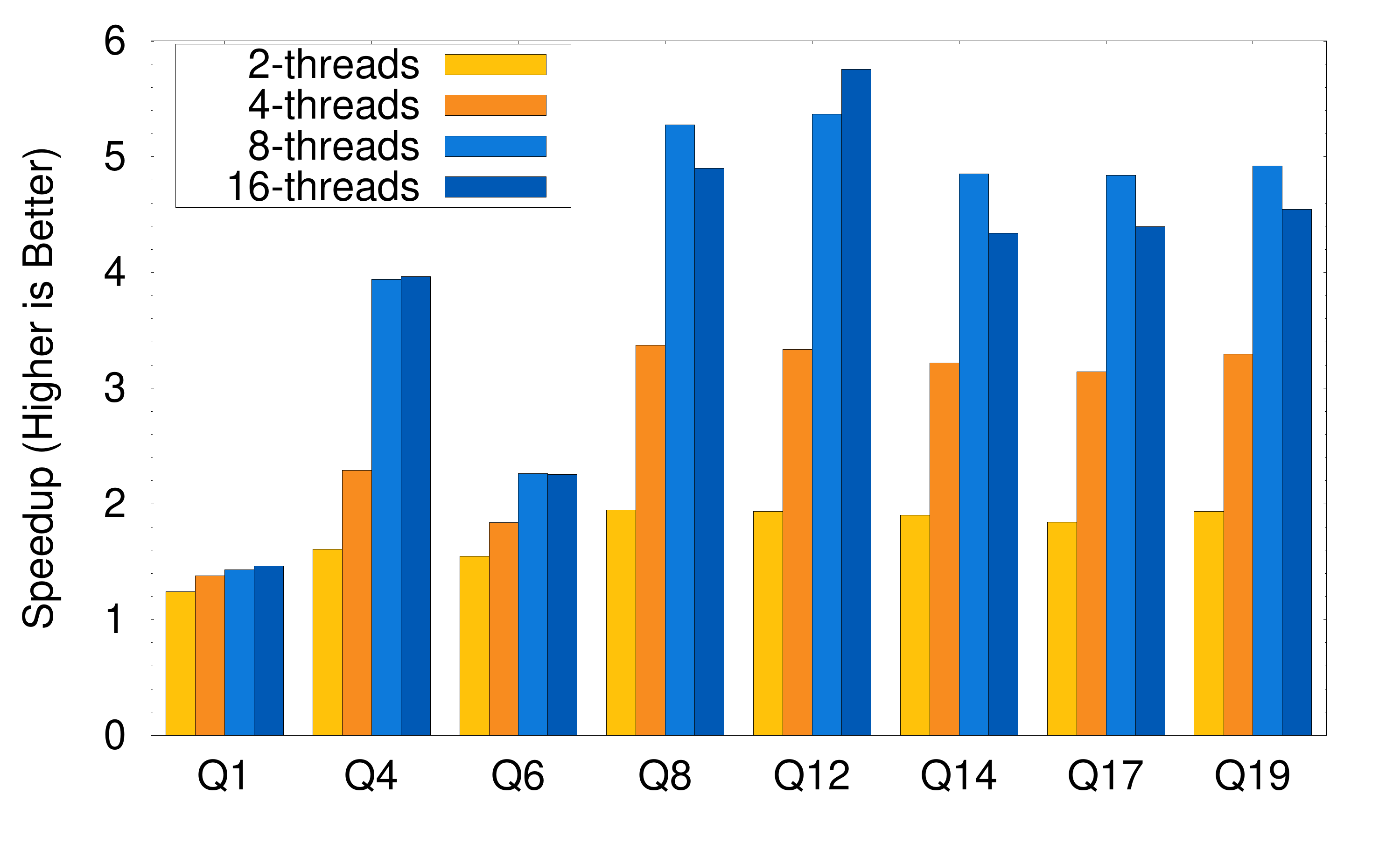}
        \vspace*{-0.2in}
        \caption{\LHEDB{}'s Multi-threaded Performance. It achieves a speedup of 1.74\x{}, 2.73\x{}, and 4.11\x{} with 2, 4, and 8 threads.}
        \label{fig:tpch_multithread}
    \end{minipage}
    \vspace*{-0.1in}
\end{figure}


Figure~\ref{fig:microbenchmark} visualizes these improvements as speedup ratios, showing \LHEDB{} consistently outperforms both baselines across all operations. The dramatic speedups for equality-based operations (3,148\x{} over \BASELINE{} and  41\x{} over ArcEDB) highlight the efficiency of our arithmetic-only approach.

\subsubsection{TPC-H Query Performance}
\begin{figure}[t]
    \centering
    \includegraphics[width=0.5\textwidth]{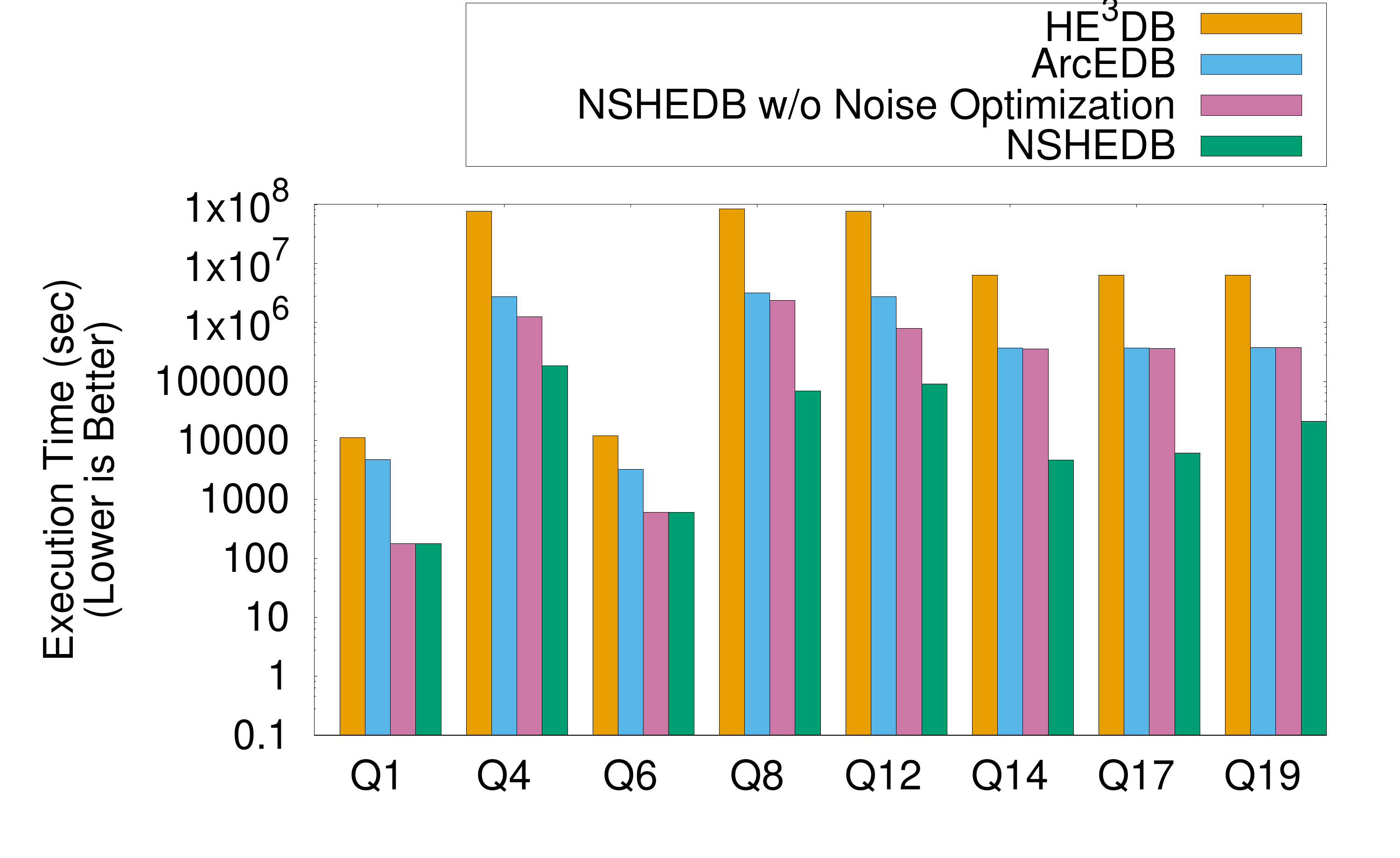}
    \vspace*{-0.1in}
    \caption{Execution time (seconds) for TPC-H queries on 32K records comparing \BASELINE{}, ArcEDB, and \LHEDB{} with and without noise optimization.}
    \label{fig:result}
    \vspace*{-0.1in}
\end{figure}

Figure~\ref{fig:result} presents a single-threaded performance comparison of \LHEDB{} against \BASELINE{} and ArcEDB across various TPC-H queries. Unlike \BASELINE{} and ArcEDB, which rely on transciphering between bit-level and arithmetic HE operations, \LHEDB{} exclusively employs arithmetic HE, eliminating the overhead of costly HE scheme conversions.

The figure shows two configurations of \LHEDB{}: without and with noise-sensitive optimizations. Even without optimization, \LHEDB{} significantly outperforms both baselines. For example, Q1 completes in 477 seconds compared to 14,454 seconds for \BASELINE{} ($30\times$ speedup) and 4,748 seconds for ArcEDB (10$\times$ speedup). Across all queries, \LHEDB{} without optimization achieves speedups between 17$times$ and 96$\times$ over \BASELINE{} (average 41$\times$).

With noise-sensitive optimizations enabled, performance improves dramatically. These optimizations restructure query plans to minimize multiplicative depth, often avoiding bootstrapping entirely. Q8 execution time drops from 8,423 seconds (\BASELINE{}) and 3,351 seconds (ArcEDB) to just 178 seconds with \LHEDB{}—speedups of 47$\times$ and 19$\times$ respectively. Overall, \LHEDB{} with optimization achieves speedups ranging from 20$times$ to 1,370$times$ over \BASELINE{} (average 659$times$) and from 5$times$ to 78$times$ over ArcEDB (average 34$times$).

Table~\ref{tab:q6-breakdown} breaks down Q6 execution time to reveal the source of these improvements. Bootstrapping dominates in HE-based systems, consuming 97.5\% of execution time in \BASELINE{} (11,509s out of 11,802s) and 84.5\% in ArcEDB (2,753s out of 3,257s). In stark contrast, \LHEDB{} eliminates bootstrapping entirely through noise-aware planning, completing the same query in just 590 seconds with only filtering and minimal aggregation overhead.

\begin{table}[h]
\centering
\small
\begin{tabular}{lrrrrr}
\toprule
\textbf{System} & \textbf{Boot.} & \textbf{Filter} & \textbf{Conv.} & \textbf{Agg.} & \textbf{Total} \\
\midrule
HE³DB & 11,509 & 251 & 42 & 0.01 & 11,802 \\
      & (97.5\%) & (2.1\%) & (0.4\%) & (<0.01\%) & \\
ArcEDB & 2,753 & 430 & 74 & 0.21 & 3,257 \\
       & (84.5\%) & (13.2\%) & (2.3\%) & (<0.01\%) & \\
NSHEDB & 0 & 589 & 0 & 1.41 & 590 \\
       & (0\%) & (99.8\%) & (0\%) & (0.2\%) & \\
\bottomrule
\end{tabular}
\caption{Execution time breakdown (seconds) for TPC-H Q6 on 32K records.}
\label{tab:q6-breakdown}
\end{table}

\subsubsection{Scalability Within Packing Limits}

Table~\ref{tab:packing-scalability} demonstrates how packing affects scaling for Q6 with varying row counts up to the packing limit ($S=32,768$):

\begin{table}[h]
\centering
\caption{Q6 execution time scaling within packing limits}
\label{tab:packing-scalability}
\small
\begin{tabular}{lccc}
\toprule
\textbf{Rows} & \textbf{\BASELINE{} (s)} & \textbf{NSHEDB (s)} & \textbf{Speedup} \\
\midrule
4K & 1,541 & 590 & 2.6$\times$ \\
8K & 3,044 & 590 & 5.2$\times$ \\
16K & 6,514 & 590 & 11.0$\times$ \\
32K & 11,802 & 590 & 20.0$\times$ \\
\bottomrule
\end{tabular}
\end{table}

\LHEDB{}'s runtime remains constant at 590 seconds for all sizes $\leq S$ (fitting within one ciphertext), while HE3DB scales linearly with data size. This demonstrates packing's efficiency: all operations within a single ciphertext have the same cost regardless of how many slots are filled. Beyond $S$ rows, \LHEDB{} would exhibit stepwise growth—requiring 2 ciphertexts for $S+1$ to $2S$ rows, 3 for $2S+1$ to $3S$ rows, and so on.

\subsection{Multi-Threaded Performance}
\label{sec:multi-thread}


\LHEDB{} exploits parallelism at multiple levels. At the data level, batch encoding enables simultaneous operations on multiple data points per ciphertext. At the query execution level, we applied OpenMP to assign independent predicates to threads, demonstrating \LHEDB{}'s parallel potential despite being orthogonal to our main contributions.

Figure~\ref{fig:tpch_multithread} illustrates performance scaling with increasing thread counts, achieving speedups of 1.74\x{}, 2.73\x{}, and 4.11\x{} with 2, 4, and 8 threads, respectively. Scaling is not linear due to three factors:

\textbf{Limited parallelism:} Queries with few predicates (e.g., Q1, Q6) leave threads idle beyond 4 cores.

\textbf{Bandwidth saturation:} Inequality checks generate intermediate ciphertexts and require lookup table accesses (BSGS optimization), creating memory pressure. Equality tests use simple squaring with minimal overhead.

\textbf{Operation dependence:} Equality-heavy queries (Q4, Q12) scale better than inequality-heavy ones (Q8, Q19) due to lower memory pressure. Joins outscale filters despite similar complexity, using more equality checks.

Future optimizations could explore memory-efficient HE schemes and predicate-aware scheduling to improve scalability beyond eight threads.

\subsection{Memory and Storage Efficiency}
\label{sec:datasize}

\begin{figure}[th]
    \begin{tabular}{cc}
    \hspace{-0.2in}
    \includegraphics[width=0.45\columnwidth]{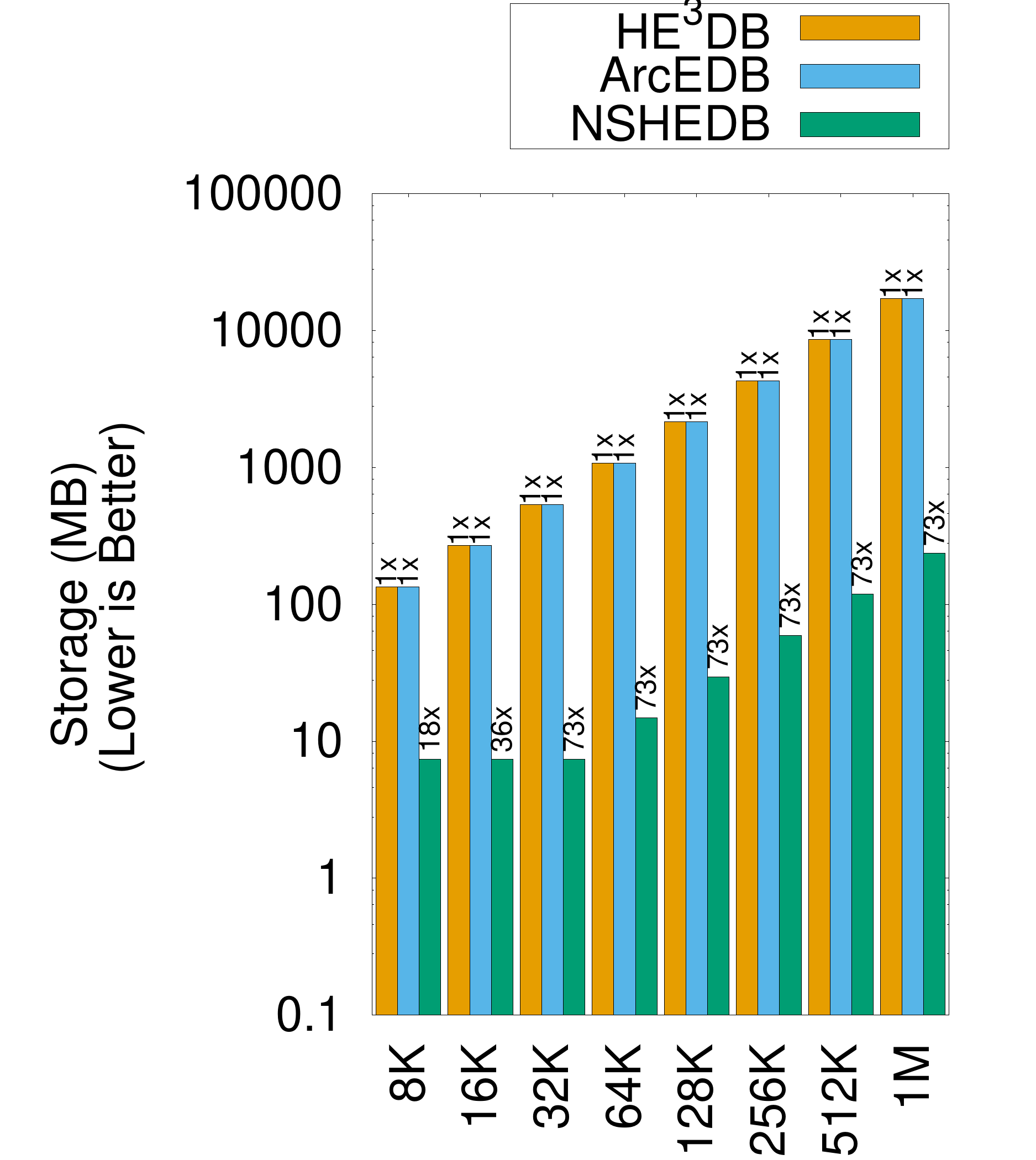} & 
    \hspace{-0.1in}\includegraphics[width=0.45\columnwidth]{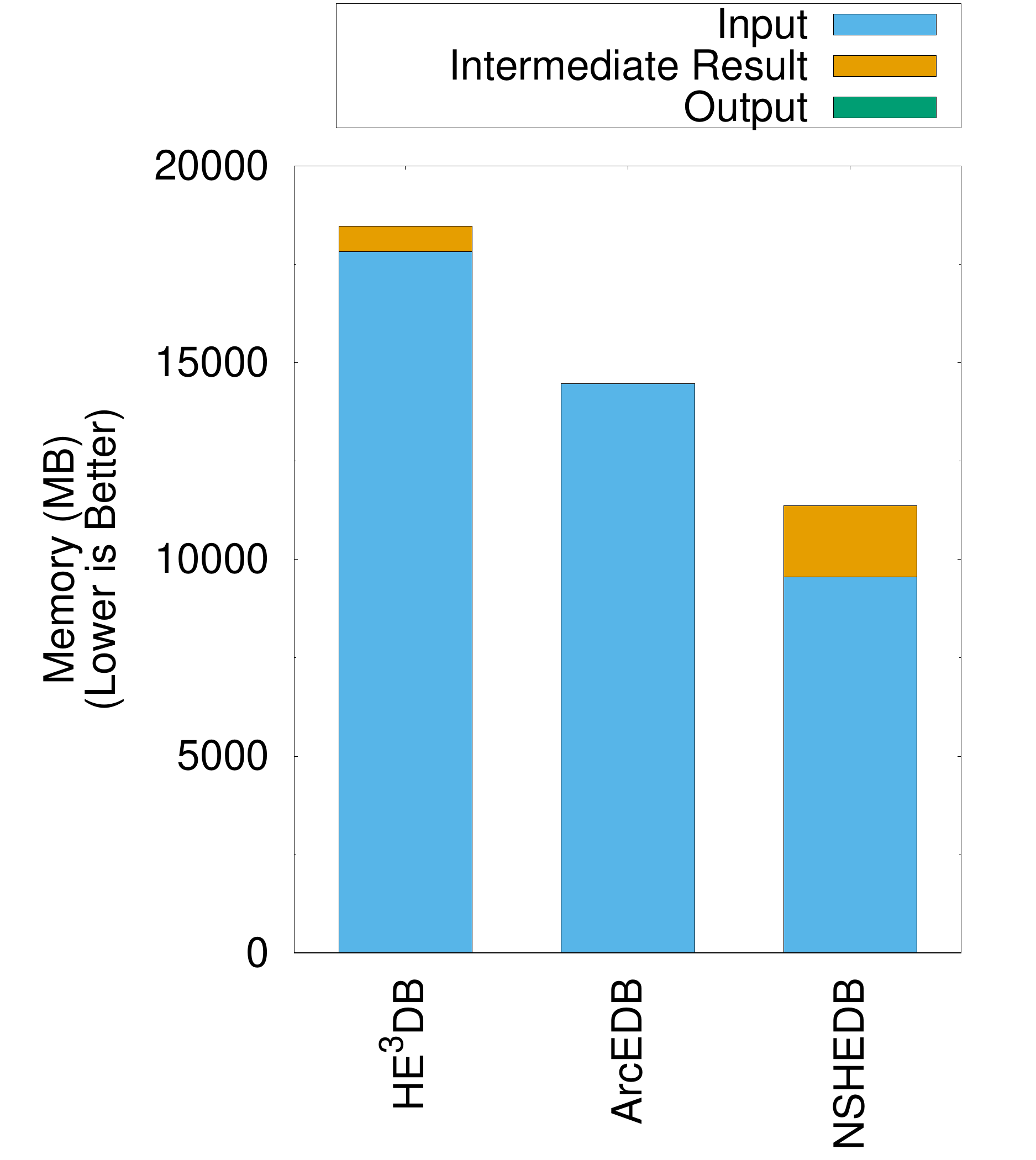} \\
    \hspace{-0.1in}(a) & \hspace{-0.1in}(b) \\
    \end{tabular}
    \vspace*{-0.1in}
    \caption{Storage and memory efficiency of \LHEDB{}. (a) Storage overhead across different record counts. \LHEDB{} reduces storage overhead by 73\x{}, requiring only 1.4\% of the space used by \BASELINE{} and ArcEDB. (b) Breakdown of in-memory object types. \LHEDB{} reduces memory overhead by up to 1.6\x{} compared to the other systems.}
    \label{fig:memory}
\end{figure}

\LHEDB{} optimizes memory and storage efficiency through improved data encoding schemes and reduced runtime memory usage. Figure~\ref{fig:memory}(a) compares the storage footprint of \LHEDB{} with \BASELINE{} and ArcEDB. To store 32K rows (16-bit integers), \LHEDB{} requires only 7.4 MB, whereas \BASELINE{} and ArcEDB require 537 MB—73\x{} more than \LHEDB{}. 

Figure~\ref{fig:memory}(b) presents a breakdown of memory usage during Q6 execution. \LHEDB{} reduces overall memory overhead by 1.6\x{} despite slightly higher intermediate memory usage. This reduction stems from two key factors:
1) Bit-level storage incurs significantly higher overhead, expanding the total memory footprint.
2) \BASELINE{} and ArcEDB require secret keys for both TFHE and CKKS, as well as CKKS scheme integration with TFHE, along with additional context data for both HE schemes. By eliminating transciphering and maintaining a more efficient memory footprint, \LHEDB{} enables more scalable and practical homomorphic query execution.

\section{Other Related Work}
\label{sec:related_work}

\subsection{Parallel Query Optimizations}

\LHEDB{}'s query optimization shares similarities with existing parallel optimization techniques but differs by focusing on reducing noise rather than raw performance. 

Existing works~\cite{ahmed2014snowstorms} and \cite{wu2011query} introduce bushy-tree-based optimizations for \textsf{JOIN} operations, aiming to reduce the number of MapReduce jobs and reorganize the sequence of operations. Similarly, prior research~\cite{bress2016robust} proposes a query chopping method for multi-threaded environments, where tasks are distributed across threads for concurrent processing.

However, these approaches primarily target raw data, while \LHEDB{} must contend with encrypted data and noise accumulation. As a result, \LHEDB{} introduces additional computational steps to minimize noise, a trade-off that slightly increases processing time but avoids the costly bootstrapping process, ultimately leading to better overall efficiency in a homomorphic encryption context.

\subsection{Multiplicative Depth Optimization}
\label{sec:noise_reduce}

Several works focus on minimizing multiplicative depth in homomorphic encryption, which is critical for optimizing performance. For boolean circuits, Aubry et al.~\cite{aubry2020faster} propose methods to minimize depth by rewriting sets of operations like XOR and AND, but these optimizations are specific to bit-level HE and may not translate well to word-level encryption schemes like BFV used in \LHEDB{}.

Bonte et al.~\cite{bonte2020homomorphic} present an optimized solution for homomorphic string search with constant multiplicative depth, focusing on equality checks. However, this approach is limited to specific query types and does not address more general database operations.

Other approaches, like the homomorphic sorting algorithm in \cite{ccetin2015depth}, prioritize depth optimization for sorting but do not extend beyond the sorting operation to subsequent computations. 

Some neural network frameworks in \cite{crockett2020low, lee2022low, lou2021hemet}, introduce depth-optimized methods for specific machine learning operations like convolution layers. However, these techniques focus on neural networks rather than general-purpose database systems, making them challenging to adapt for \LHEDB{}'s query processing needs.
\section{\reviewed{Conclusion}}
\label{sec:conclude}
Our work introduces \LHEDB{}, a secure query processing engine that rethinks how homomorphic encryption (HE) can be deployed in practical database systems. Rather than relying on costly transciphering or trusted hardware, \LHEDB{} uses word-level HE and introduces novel algorithms for executing common database operations entirely within the encrypted domain. It promotes a new approach to HE-based query execution by translating logical operations into efficient arithmetic equivalents, while prioritizing noise growth control over simply minimizing operation counts.

By integrating noise-aware query planning with carefully optimized encodings, \LHEDB{} avoids excessive bootstrapping and significantly improves computational efficiency. Our evaluation on TPC-H workloads shows that \LHEDB{} achieves up to 1370$\times$ speedup over state-of-the-art HE-based systems, with only 1.4\% storage overhead, and no reliance on trusted execution environments, key release, or scheme-switching.

This work demonstrates that secure, privacy-preserving query processing with HE is not only theoretically sound but also practical and performant. \LHEDB{} sets a new direction for building encrypted database systems grounded in principled cryptographic design and system-level innovation.

\section*{Acknowledgments}
The authors would like to thank the anonymous reviewers
for their helpful comments.  
This work was sponsored by the National Science Foundation (NSF) award, 
CNS-2231877, and Intel.



%

\bibliographystyle{plain}
\bibliography{paper}


\end{document}